\documentclass{article}
\usepackage{graphicx}
\usepackage{enumerate}
\usepackage{color}
\usepackage{hyperref}
\usepackage{fullpage}
\usepackage{amsmath,amssymb}
\usepackage{cite}
\usepackage{lineno}
\usepackage{authblk}
\usepackage{color}

\newcommand{\beginsupplement}{%
	\setcounter{table}{0}
	\renewcommand{\thetable}{S\arabic{table}}%
	\setcounter{figure}{0}
	\renewcommand{\thefigure}{S\arabic{figure}}%
}

\begin{document}

\bibliographystyle{unsrt}

\title{Diversity of meso-scale architecture in human and non-human connectomes}
\author[1]{Richard F. Betzel}
\author[2]{John D. Medaglia}
\author[1,3]{Danielle S. Bassett}
\affil[1]{Department of Bioengineering}
\affil[2]{Department of Psychology}
\affil[3]{Department of Electrical and Systems Engineering \newline University of Pennsylvania, Philadelphia, PA, 19104}

\maketitle

\begin{abstract}
The brain's functional diversity is reflected in the meso-scale architecture of its connectome, i.e. its division into clusters and communities of topologically-related brain regions. The dominant view, and one that is reinforced by current analysis techniques, is that communities are strictly assortative and segregated from one another, purportedly for the purpose of carrying out specialized information processing. Such a view, however, precludes the possibility of non-assortative communities that could engender a richer functional repertoire by allowing for a more complex set of inter-community interactions. Here, we use weighted stochastic blockmodels to uncover the meso-scale architecture of \emph{Drosophila}, mouse, rat, macaque, and human connectomes. We confirm that while many communities are assortative, others form core-periphery and disassortative structures, which in the human better recapitulate observed patterns of functional connectivity and in the mouse better recapitulate observed patterns of gene co-expression than other community detection techniques. We define a set of network measures for quantifying the diversity of community types in which brain regions participate. Finally, we show that diversity is peaked in control and subcortical systems in humans, and that individual differences in diversity within those systems predicts cognitive performance on Stroop and Navon tasks. In summary, our report paints a more diverse portrait of connectome meso-scale structure and demonstrates its relevance for cognitive performance.

\end{abstract}

\section*{Introduction} \label{Introduction}

Cognitive processes are thought to emerge from the coordinated activity of distributed networks of neural elements, from small-scale neuronal populations to large-scale brain areas \cite{bressler2010large, medaglia2015cognitive, mivsic2016regions}. This coordination is facilitated by the brain's network of physical, hard-wired connections -- its \emph{connectome} \cite{sporns2005human, honey2009predicting, deco2013resting, hermundstad2013structural, goni2014resting}. Accordingly, the range of cognitive processes in which a neural element participates as well as its computational capacity depends critically on its connectivity profile, i.e. its set of outgoing and incoming connections along which it transmits information to and receives information from other brain areas \cite{levy2001distributed, sporns2004motifs, fiete2010spike, hermundstad2011learning, eliasmith2012large, rajan2016recurrent}.

While individual neural elements are thought to perform local operations, their organization into motifs, circuits, and communities engenders a richer, more diverse functional repertoire \cite{sporns2000theoretical, honey2007network, senden2014rich, zamora2016functional}. In particular, the connectome's communities -- which collectively comprise its so-called \emph{meso-scale structure} -- have attracted a great deal of attention (see \cite{sporns2016modular} for a recent review). Here, we define the meso-scale as the level between that of individual nodes and the network as a whole \cite{betzel2016multi}. At that scale, a network's nodes can be grouped into clusters called ``communities'', which are usually assumed to be assortative, meaning that nodes preferentially connect to nodes with similar attributes, namely membership to the same community \cite{newman2003mixing}. The resulting communities are internally dense and externally sparse, and are oftentimes described as ``nearly decomposable'', segregated, and autonomous \cite{simon1962architecture}.

The assortative community model has informed our current understanding of brain network function, perpetuating a stylized view of the brain in which segregated (i.e. assortative) communities engage in specialized information processing while a small number of highly-connected hubs integrate information across communities \cite{hagmann2008mapping, zamora2010cortical, van2013network}. This view is supported by cross-species analyses uncovering analogous structure in both human and non-human connectome data \cite{harriger2012rich, towlson2013rich, de2013rich, shih2015connectomics, van2011rich}, suggesting that assortative communities may be an evolutionarily-conserved architectural feature. In addition, computational models have shown that networks in which assortative communities are interlinked by hubs facilitate complex neural dynamics, including that necessary for the spread of information \cite{mivsic2015cooperative, zamora2016functional, worrell2017signal}.

\begin{figure*}[t]
	\begin{center}
		\centerline{\includegraphics[width=0.8\textwidth]{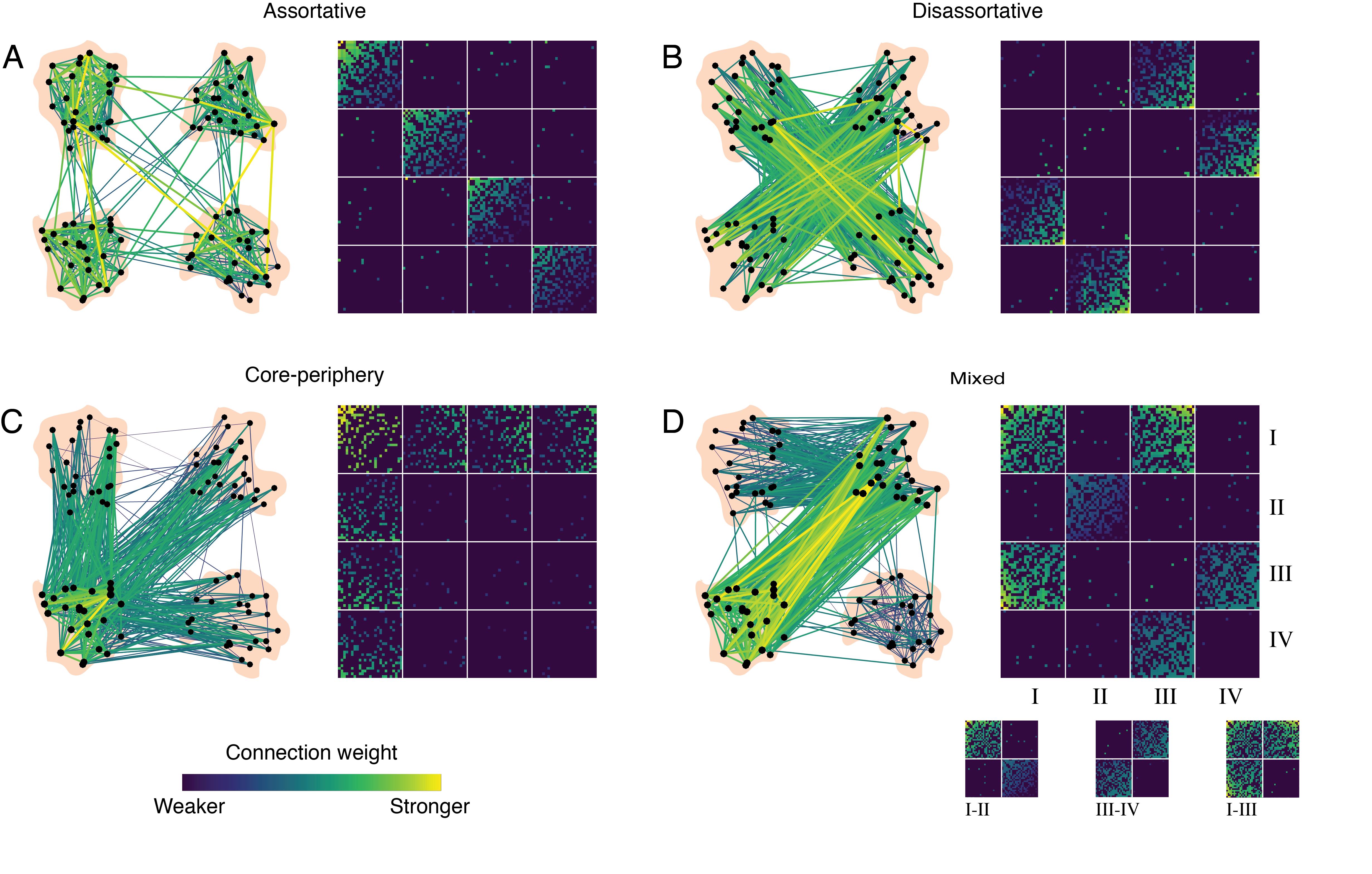}}
		\caption{\textbf{Community structure types.} Networks can exhibit different types of meso-scale structure. (\emph{A}) Assortative communities are sub-networks whose internal density of connections exceeds their external density. (\emph{B}) Disassortative (multi-partite) communities are sub-networks where connections are made preferentially between communities so that communities' external density exceed their internal density. (\emph{C}) Core-periphery organization consists of a central core that is connected to the rest of the network and then peripheral nodes that connect to the core but not to one another. (\emph{D}) These meso-scale structures can be present simultaneously in the same network. For example, communities \emph{I-II} interact assortatively, \emph{III-IV} interact disassortatively, while \emph{I-III} interact as a core and periphery.} \label{exampleCommunityTypes}
	\end{center}
\end{figure*}

While this perspective has proven useful, it has a number of drawbacks, of which we focus on two. First, it makes the strong assumption that connectome meso-scale architecture is strictly assortative (Fig.~\ref{exampleCommunityTypes}A). This assumption stems in part from the algorithms used to detect communities, the most popular of which seek internally dense and externally sparse sub-networks, and therefore preclude the possibility of detecting non-assortative structure \cite{newman2004finding, rosvall2008maps}. It is unclear, then, whether the assortative communities uncovered using these algorithms represent an accurate picture of connectome meso-scale structure or whether they reflect the assumptions and limitations of the algorithms themselves.

Secondly, this perspective forces a binary and somewhat arbitrary distinction between architectural features of connectomes that support segregated \emph{versus} integrated brain function (communities and hubs, respectively), without allowing the possibility of a richer, diverse perspective on the types of computations that may be necessary to support cognition. In general, meso-scale structure can be non-assortative. For instance, \emph{dis}assortative structure (Fig.~\ref{exampleCommunityTypes}B) refers to communities that preferentially avoid forming connections to the other members of their own community \cite{clauset2008hierarchical}. Core-periphery organization (Fig.~\ref{exampleCommunityTypes}C), on the other hand refers to the presence of a dense, central core community that links to a peripheral set of communities that are sparsely connected among themselves \cite{borgatti2000models,csermely2013structure}.

Both disassortative and core-periphery organization represent meso-scale architectures that allow communities to interact strongly with one another. Accordingly, networks that exhibit either type have the capacity to perform a different set of functions than a network composed of purely assortative communities. For example, a core community has the capacity to exert influence over and control the more peripheral communities \cite{gu2015controllability, betzel2016optimally}. Similarly, disassortative communities lack internal connections, suggesting that communities can influence one another but may lack mechanisms for self-regulation \cite{clauset2008hierarchical}.

Most importantly, a network can express multiple classes of meso-scale structure simultaneously, allowing for certain subsets of nodes to interact assortatively, disassortatively, and as cores and peripheries (Fig.~\ref{exampleCommunityTypes}D). Here, we hypothesize that diverse meso-scale structure allows a network to engage in a wider functional repertoire, in general. As a secondary hypothesis, we suppose that inter-subject variability in diversity is predictive of variation in cognitive performance on tasks that require coordination among a diverse set of cognitive systems.

\begin{figure*}[t]
	\begin{center}
		\centerline{\includegraphics[width=0.75\textwidth]{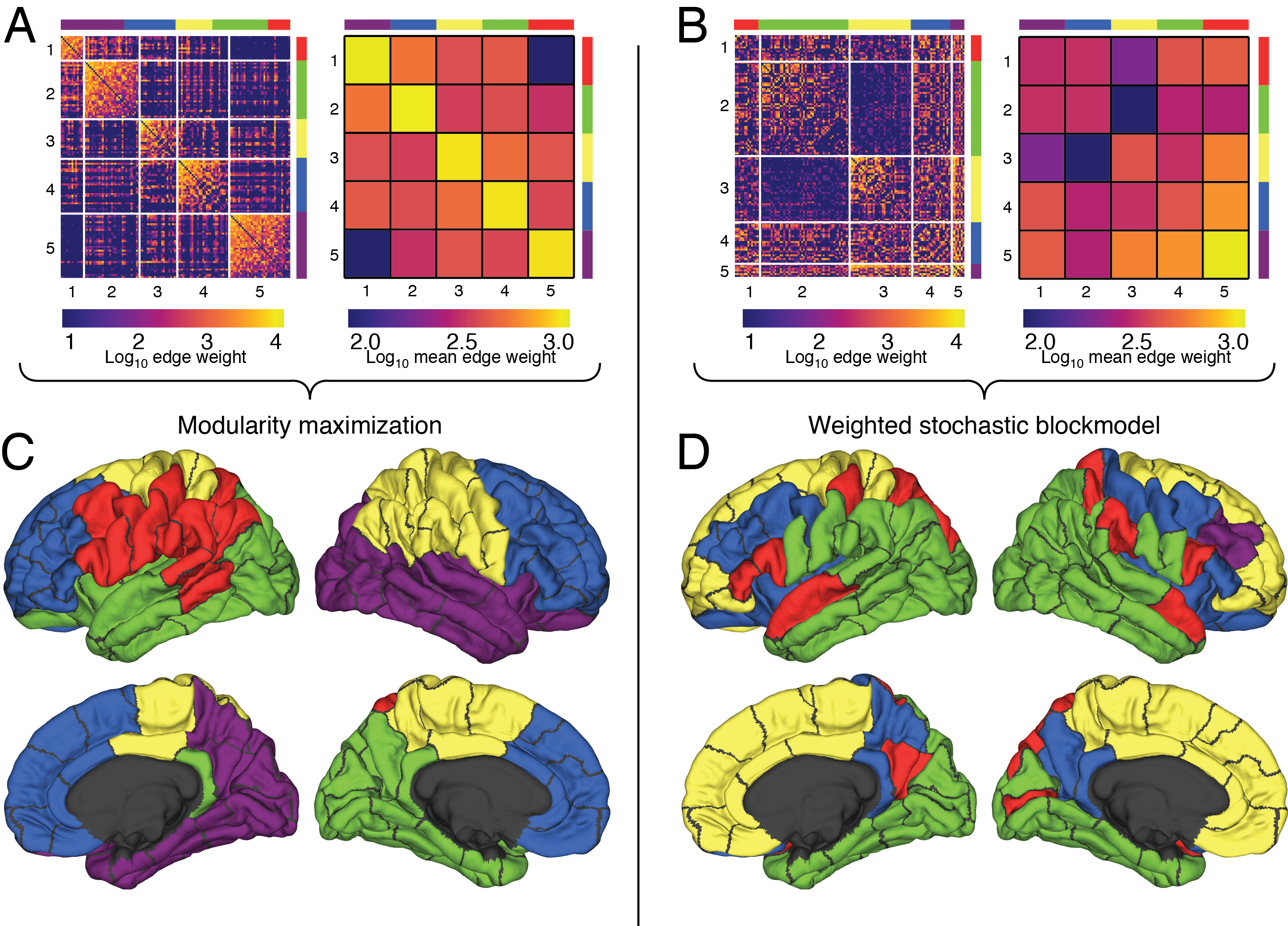}}
		\caption{\textbf{Example WSBM and $Q_{max}$ communities.} (\emph{A}) Human connectome network ordered by community partitions detected using $Q_{max}$ and (\emph{B}) the WSBM. Both examples are shown with the number of communities fixed at $K = 5$. The color of matrix elements for the \emph{left} sub-panels represents log-transformed edge weights while the color of elements for the \emph{right} sub-panels represents the log-transformed mean within- and between-community edge weights.	Panels (\emph{C}) and (\emph{D}) depict the spatial distributions of those same partitions.} \label{basicNetworkComparison_human}
	\end{center}
\end{figure*}

In this report, we address these hypotheses by using the weighted stochastic blockmodel to uncover the meso-scale architecture of both human and non-human connectome data \cite{aicher2013adapting, aicher2014learning}. We show that, in addition to assortative communities described in previous reports, connectomes show evidence of non-assortative structure, including subsets of communities that interact disassortatively and others that form cores and peripheries. Next, we define a node-level diversity index that quantifies the extent to which individual neural elements participate in communities of all classes. We show that in humans, this index is peaked in regions associated with cognitive control and in sub-cortical areas, suggesting that traditionally defined cortico-subcortical circuits that support cognitive control are positioned to participate in a wide range of meso-scale processes. Finally, we show that diversity of connections in these same systems predicts cognitive individual differences in performance on two cognitive control tasks: namely, the Stroop \cite{van2005separating} and Navon \cite{navon1977forest} tasks.

\section*{Results} \label{Results}

We fit the weighted stochastic blockmodel (WSBM) to connectome data acquired from humans, resulting in partitions of each networks' nodes into $K$ communities (See \textbf{Materials and Methods} for details on connectome reconstruction and mathematical underpinnings of the WSBM). To ensure that our results are not biased by a specific reconstruction technique, we also fit the WSBM to non-human connectome data including mouse, rat, macaque, and \emph{Drosphila} (See \textbf{Supplementary Materials} for more details on non-human data; Fig.~\ref{basicNetworkComparisonNonHuman1}). Importantly, the WSBM is flexible and not limited to the detection of assortative communities, but can also detect disassortative communities and core-periphery organziation. As a point of comparison, we also obtained partitions using modularity maximization ($Q_{max}$) \cite{newman2004finding}, a community detection technique that is currently widely used in network neuroscience \cite{bassett2017network}. Unlike the flexibility of the WSBM, $Q_{max}$ is designed to detect only assortative community structure.

Rather than focus on partitions of the network into a specific number of communities based on a single partition, we analyzed the types of communities identified as the number of communitie sought ranged over $K = 2, \ldots , 10$. As demonstrated in Figs.~\ref{communitySimilarity3}, \ref{fcMat_human} -- \ref{behavioralAnalysis}, this approach enabled us to demonstrate that our results were robust to reasonable variation in $K$. In the interest of parsimony, Fig.~\ref{basicNetworkComparison_human} and Fig.~\ref{maximallyAssortativeSetHuman} depict results using an intermediate number of communities, namely $K = 5$.

\subsection*{Connectomes support diverse meso-scale architecture}

It is widely believed that segregated information processing and, hence, assortative meso-scale structure is one of the principles driving the organization of connectomes \cite{van2013anatomical, sporns2016modular} (Fig.~\ref{basicNetworkComparison_human}A,C). Alternative classes of meso-scale structure, including those that theoretically support diverse neural computations, are rarely considered. This point of view gets further reinforced by analysis techniques that favor the detection of assortative structures. Few studies, however, have explored what effect alternative methods for uncovering mesoscale structure might have on our understanding of connectome organization and function. Here, we use the more flexible WSBM to detect the meso-scale structure of connectomes (Fig.~\ref{basicNetworkComparison_human}B,D). We compare the WSBM partitions to those detected using $Q_{max}$ at three distinct topological scales: namely whole-brain, community, and region (node) levels.

\begin{figure*}[t]
	\begin{center}
		\centerline{\includegraphics[width=1\textwidth]{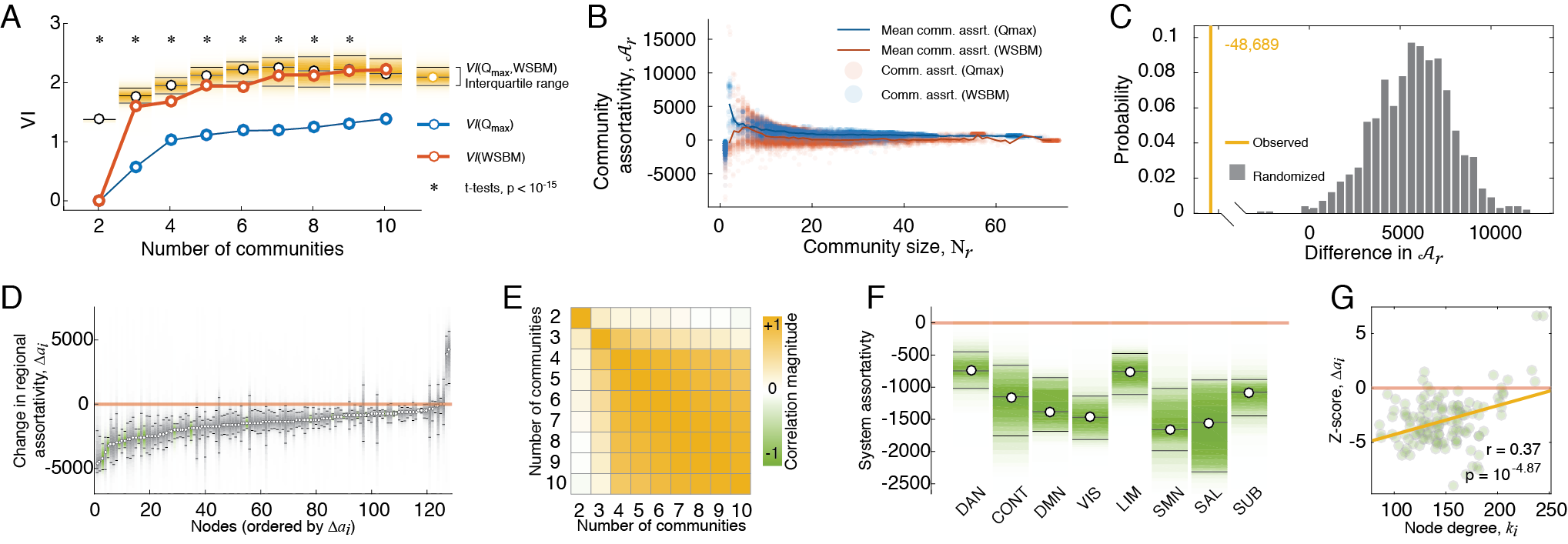}}
		\caption{\textbf{Modularity maximization and the weighted stochastic blockmodel uncover fundamentally different architectural signatures.} (\emph{A}) Variation of information within and across community detection techniques ($Q_max$ and WSBM) demonstrating greater dissimilarity of partitions between techniques than within techniques. (\emph{B}) Community assortativity, $\mathcal{A}$, as a function of community size, $N$, demonstrating that $Q_{max}$ communities, on average, are more assortative. (\emph{C}) Comparison of statistic derived using functional data analysis (yellow line) with that expected in a null distribution. Specifically, we generated a statistic by performing a pointwise subtraction and summation of the curves $\bar{\mathcal{A}}(N)$ obtained for the WSBM and $Q_{max}$. The value of this statistic quantifies the difference between mean community assortativity across communities of all sizes and is negative when communities detected using $Q_{max}$ are more assortative than WSBMs. We compared this statistic against a null distribution obtained from a null model wherein we perserved the number and size of communities in a given partition but permute nodes' assignments uniformly and randomly (1000 repetitions). (\emph{D}) Changes in regional assortativity, $\Delta a_i$, when considering WSBM \emph{versus} $Q_{max}$ partitions, ordered by greatest to least decrease. Note that the majority of regions decrease assortativity in the partitions estimated from WSBM compared to those estimated from $Q_{max}$ (i.e. $\Delta a_i < 0$). (\emph{E}) Correlation of regional assortativity while varying the number of communities from $K = 2, \ldots ,10$. Note the high consistency for $K \ge 4$. (\emph{F}) Regional assortativity scores grouped by cognitive systems. DAN=dorsal attention, CONT=cognitive control, DMN=default mode, VIS=visual, LIM=limbic, SMN=somatomotor, SAL=salience, SUB=subcortical. (\emph{G}) Regional assortativity (corrected for degree through standardization procedure) as a function of node degree, $k_i$.} \label{communitySimilarity3}
	\end{center}
\end{figure*}

To assess whole-brain similarity of WSBM and $Q_{max}$ partitions, we computed pairwise variation of information (VI) which is nonzero for dissimilar partitions and zero when they are identical \cite{meilua2003comparing} (See \textbf{Materials and Methods} for mathematical details; Fig.~\ref{communitySimilarity3}A for results). In order to make the comparison as fair as possible, we controlled for the number of communities, $K$, and only computed VI among partitions that resulted in the same number of communites. This procedure resulted in a series of within- and between- technique VI scores as a function of $K$. At each $K$, we computed one-tailed $t$-tests to assess whether the mean within-technique dissimilarity was smaller than the between-technique dissimilarity. We observed that from $K = 2, \ldots , 9$, both the WBSM and $Q_{max}$ uncovered partitions that were self-consistent yet distinct from one another (maximum $p < 10^{-15}$). This observation was consistent across the non-human connectome data as well (Fig.~\ref{communitySimilarity4}). These results confirm that the WSBM and $Q_{max}$ generate statistically different estimates of connectome community structure.

Next, we wished to confirm that the WSBM uncovered non-assortative communities, which in theory could support the observed diversity in cognitive functions. To test this hypothesis, we computed for each community $r$, its size, $N_r$, and assortativity score, $\mathcal{A}_r$, which measured its internal density of connections less its maximum density of connections to any other community (See \textbf{Materials and Methods}). We then aggregated all detected communities and computed the mean assortativity score as a function of community size, $\bar{\mathcal{A}}(N)$ (Fig.~\ref{communitySimilarity3}B). These procedures were performed separately for the WSBM and $Q_{max}$. We compared these curves using functional data analysis, which is a set of statistical tools for comparing continuous curves \cite{ramsay2002applied, ramsay2006functional}. We found that the observed scores were smaller than those obtained under the null model ($p < 10^{-3}$), confirming that WSBM communities tend to be less assortative than $Q_{max}$ (Fig.~\ref{communitySimilarity3}C). Again, these findings are consistent across connectome data obtained from all species (Fig.~\ref{communityAssortativity4}).

Finally, because the functional roles of brain regions also depend on local connections to their own and other communities, we tested whether our understanding of these roles changed when we considered WSBM communities rather than those uncovered using $Q_{max}$. Specifically, we examined how regional assortativity scores, a node-level metric analogous to community assortativity, differed given the WSBM \emph{versus} $Q_{max}$ partitions (See \textbf{Materials and Methods}; Fig.~\ref{communitySimilarity3}D-E). We found regional assortativity decreased for most nodes. Aggregating differences by functional systems \cite{yeo2011organization} (See \textbf{Materials and Methods}), we found that the greatest decrements were concentrated within visual and somatomotor systems (Fig.~\ref{communitySimilarity3}F). Interestingly, decreased regional assortativity was also correlated with node degree (the number of connections a node makes), with low-degree nodes exhibiting greater decreases compared to high-degree nodes ($r = 0.37$, $p < 10^{-4}$; Fig.~\ref{communitySimilarity3}G).

In summary, these findings confirm that the weighted stochastic block model and modularity maximization uncover communities of fundamentally different nature. Among the most profound differences is the assortativity of detected communities, with the WSBM consistently detecting less assortative communities (i.e. less segregated and more integrated) than $Q_{max}$. This is relevant from a psychological and neuroscientific point of view: $Q_{max}$ provides a picture of the connectome that is in line with the notion that cognitive processes are modular \cite{fodor1983modularity}, carried out by segregated populations of neurons and groups of brain regions \cite{sporns2016modular}. By constrast, WSBM paints a richer picture of the structural support for a potentially more diverse set of dynamics; the reduction in assortativity (and hence segregation) implies increased interdependence of communities on one another.

\subsection*{Many (but not all) communities are assortative}

In the previous section, we provided evidence that connectomes exhibit diverse, non-assortative communities, suggesting the capacity for an equally diverse cognitive repertoire. This finding, however, runs counter to the dominant narrative surrounding brain network function, namely that information processing is carried out by specialized, assortative communities. An important question, then, is whether the reduction in assortativity and increase in community diversity described in the previous section are driven by a small subset of non-assortative communities (so that most communities are still assortative) or whether all communities uniformly decrease in assortativity. From a purely theoretical standpoint, there are myriad reasons why it might be advantageous for a network to retain some assortative communities. For example, in the interest of promoting functional robustness and evolvability \cite{kirschner1998evolvability}. Neuroscientifically, this is an important question because it further informs our understanding of the anatomical specificity of network topologies that underpin and support cognitive function.

\begin{figure*}[t]
	\begin{center}
		\centerline{\includegraphics[width=1\textwidth]{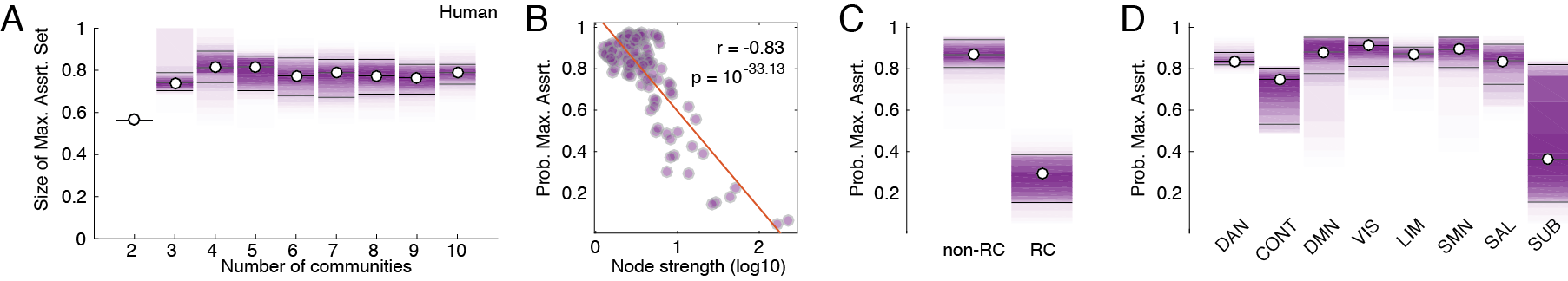}}
		\caption{\textbf{Maximally assortative set.} (\emph{A}) Fraction of brain regions comprising the maximally assortative set as a function of the number of communities. (\emph{B}) High strength nodes are less likely to participate in the maximally assortative set. (\emph{C}) As a consequence, the maximally assortative set is comprised mostly of non-rich club brain regions; rich club = RC. (\emph{D}) At the system level, control and subcortical networks are the least likely to participate in the maximally assortative set. DAN=dorsal attention, CONT=cognitive control, DMN=default mode, VIS=visual, LIM=limbic, SMN=somatomotor, SAL=salience, SUB=subcortical.} \label{maximallyAssortativeSetHuman}
	\end{center}
\end{figure*}

To test this, we uncovered the maximally assortative set of communities for each WSBM partition, which comprises the largest set of communities (in terms of the number of nodes included in those communities) whose minimum within-community density of connections exceeds its maximum between-community density (See \textbf{Materials and Methods}). We then estimated how frequently, on average, each brain region participated in this set. We found that as we varied the number of communities from $K = 2$ to $K = 10$, the maximally assortative set comprised $75\pm13$ percent of all nodes (Fig.~\ref{maximallyAssortativeSetHuman}A). In general, the maximally assortative set was also comprised of low strength, non-rich club nodes (Fig.~\ref{maximallyAssortativeSetHuman}B,C). Breaking down inclusion in this set by cognitive system, we found that control and subcortical systems were the least likely to be included in the maximally assortative set compared to the other systems (Fig.~\ref{maximallyAssortativeSetHuman}D).

In summary, these findings confirm that while the WSBM tends to detect less assortative communities than $Q_{max}$, there nonetheless exists a backbone of highly assortative communities that, as a group, exhibit the ubiquitous internally-dense, externally-sparse connection density. This collection of communities, which largely excludes the brain's highly connected regions, therefore has the capacity to perform segregated information processing.

\subsection*{Functional relevance of the WSBM}

To this point, we have used the WSBM to demonstrate that connectomes exhibit diverse, non-assortative meso-scale structure. We contrasted the outputs of each algorithm and showed that such differences are manifest at all topological scales, ranging from the partition as a whole to individual network nodes \cite{betzel2016multi}. Based on these observations, we hypothesize that partitions detected by the WSBMs better account for the brain's functional network organization than those detected by $Q_{max}$, suggesting that the richer, non-assortative communities are closer to the brain's canonical architecture.

To test this hypothesis, we compare the communities detected using both the WSBM and $Q_{max}$ with a group-representative estimate of the brain's functional connectivity (FC) network estimated from resting state fMRI data (See \textbf{Materials and Methods} for more details on FC reconstruction from BOLD signals). While there are many approaches for establishing a relationship between brain structure and function \cite{damoiseaux2009greater, abdelnour2014network, stam2016relation}, perhaps the simplest is to compare the average within- and between-community densities of functional connections. Intuitively, functionally-related brain regions are linked by strong functional connections. If a community does a good job identifying sets of such regions, then the within-community density of functional connections should be greater than the between-community density.

\begin{figure*}[t]
	\begin{center}
		\centerline{\includegraphics[width=0.8\textwidth]{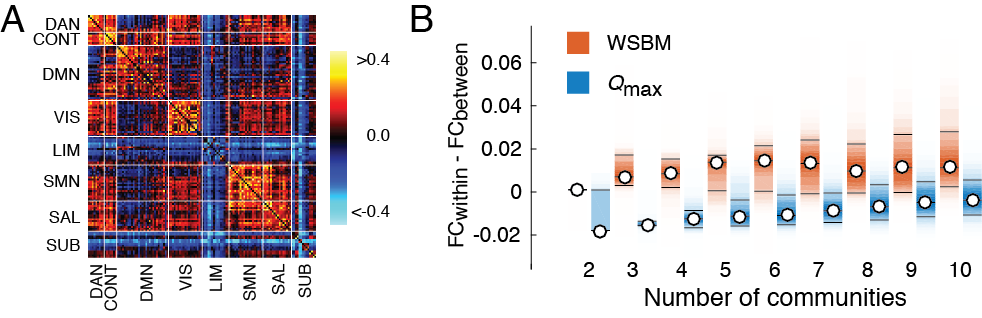}}
		\caption{\textbf{Communities estimated from the weighted stochastic block model are more functionally segregated than communities estimated from modularity maximization.} (\emph{A}) Functional connectivity (FC) matrix ordered by functional system: DAN=dorsal attention, CONT=cognitive control, DMN=default mode, VIS=visual, LIM=limbic, SMN=somatomotor, SAL=salience, SUB=subcortical. (\emph{B}) Difference between within- and between-community FC for the WSBM (orange) and $Q_{max}$ (blue). Each box plot depicts the variance over partitions detected using either the WSBM or $Q_{max}$.} \label{fcMat_human}
	\end{center}
\end{figure*}

To test whether this was the case, we imposed partitions obtained from the WSBM and $Q_{max}$ applied to the structural connectome onto the FC matrix and computed the difference of within- and between-community FC density. We found that over a range $K = 2, \ldots, 10$, the WSBM consistently uncovered communities whose internal FC density exceeded their between-community density (Fig.~\ref{fcMat_human}A). Moreover, the difference of within- and between-community FC density was greater using the WSBM communities than with $Q_{max}$ communities ($t$-tests, $p < 0.01$; Fig.~\ref{fcMat_human}B), suggesting that the WSBM communities capture functional relationships among brain regions. We also report consistent findings when we apply the same methodology to correlated gene expression patterns for the mouse connectome (Fig.~\ref{geneMat}).

\subsection*{Community morphospace reveals rules for between-community interactions}

To this point we have shown that the brain exhibits rich meso-scale structure that can be uncovered using the WSBM, and that offers a better explanation for human brain function and mouse genetic expression than that provided by assortative communities alone. In this section, we seek a fundamental understanding of the exact nature of that mesoscale architecture, and therefore ask the question: ``How do interactions among pairs of communities combine to generate assortative and non-assortative meso-scale architecture?'' To address this question, we focus our analysis onto the interactions among pairs (dyads) of communities. Community dyads represent the building blocks of a network's meso-scale structure, and can be combined in different configurations and proportions to engender larger, more complex functional circuits. We investigate community interactions using a theoretical morphospace analysis \cite{mcghee2006geometry}, a technique recently adapted to the study of complex networks \cite{avena2014using, avena2015network, goni2013exploring}.

A morphospace is a hyperspace whose axes are features of a particular class of organism or system. In the case of complex networks, axes usually are defined to be topological properties of a network, e.g. efficiency, wiring cost, modularity, \cite{goni2013exploring} or the parameters of generative network models \cite{vertes2012simple, betzel2016generative}. Once the axes are defined, any observed network can be represented in the morphospace as a point whose location is defined by that network's particular combination of features. In general, morphospaces are not uniformly populated. Evolutionary and functional constraints render some regions more favorable (and hence more densely populated) than others, and by studying the density of points throughout the morphospace one can better understand how those constraints influence the structure of a network.

\begin{figure*}[t]
	\begin{center}
		\centerline{\includegraphics[width=1\textwidth]{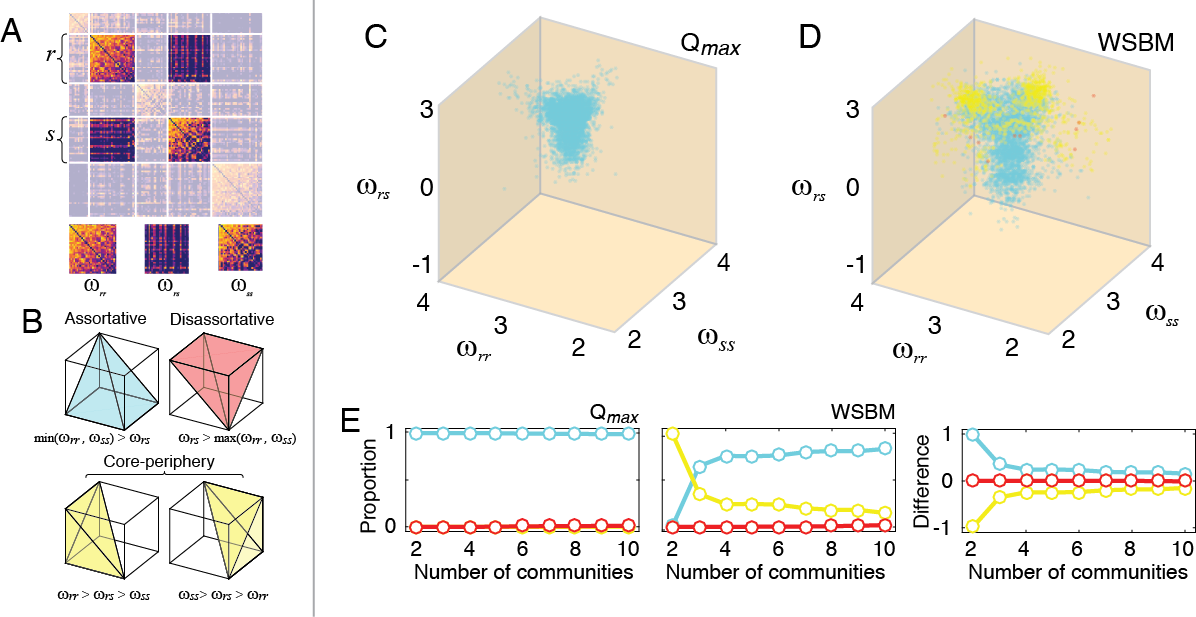}}
		\caption{\textbf{A rich community morphospace.} (\emph{A}) A community motif is constructed as the average over blocks of the connectivity matrix. Here, we show blocks within and between two communities, labeled \emph{r} and \emph{s}. (\emph{B}) Given within- and between-community connection densities, it is possible to classify each pair of communities into one of three motifs: assortative, disassortative, or core-periphery. (\emph{C},\emph{D}) All pairs of communities placed in a network morphospace and colored by their motif type (note: axes are log-scaled). (\emph{E}) The relative proportion of each motif type as a function of the number of detected communities, $K$, for $Q_{max}$ (\emph{left}), the WSBM (\emph{middle}), and their difference (\emph{right}).} \label{communityMorphospaceWeighted_newFigs}
	\end{center}
\end{figure*}

Here, rather than constructing a morphospace of networks, we construct a three-dimensional community morphospace, allowing us to investigate how interactions among pairs of communities combine to generate assortative and non-assortative meso-scale architecture. In this morphospace, each point represents a pair of communities, $\{r, s\}$, and the axes are defined to be their respective within- and between-community connection densities, $\omega_{rr}$, $\omega_{ss}$, and $\omega_{rs}$ (Fig.~\ref{communityMorphospaceWeighted_newFigs}A). These features can be used to classify the interaction of $r$ and $s$ into one of three canonical community interaction motifs: assortative, core-periphery, and disassortative (See \textbf{Materials and Methods} for details) (Fig.~\ref{communityMorphospaceWeighted_newFigs}B).

We compared morphospaces constructed based on communities detected using the WSBM and $Q_{max}$, and computed the relative proportion of each motif type (Fig.~\ref{communityMorphospaceWeighted_newFigs}C,D). Across $K = 2, \ldots , 10$, we found that $Q_{max}$ partitions resulted in, almost exclusively, assortative interactions among communities. The WSBM also favored assortative interactions, but included a significant number of core-periphery and disassortative interactions. Again using functional data analysis, we compared the relative proportion of each motif by performing a pointwise subtraction and then summation of each motif's relative proportion as a function of $K$ and aggregated the motif-specific scores to generate a statistic. This statistic measured the absolute difference in relative motif proportion as the number of communities varied from $K = 2$ to $K = 10$. We compared this statistic against a null distribution generated by randomly and uniformly permuting nodes' community assignments and recalculating motif proportions. We found that the observed difference exceeded what would be expected by chance ($p = 0.029$; 1000 permutations), indicating that the relative proportion of community motifs discovered using the WSBM was different than $Q_{max}$ and could be largely explained by increased diversity of motif types using the WSBM. Again, this result was also observed in non-human connectome data (Fig.~\ref{communityMorphospaceWeighted_nonHuman}).

\subsection*{Community motifs identify a class of diversely connected nodes}

Community motifs represent interactions among pairs of communities. We mapped motifs back to the level of individual brain regions by computing a motif participation index for each brain region, which measured the fraction of times the community that region was assigned to participated in each motif. Importantly, for the core-periphery motif, we distinguished between the ``core'' community and the ``periphery'' community. In addition, we also computed a diversity index: an entropy over the motif participation distribution (See \textbf{Materials and Methods}). A region that participated largely in one motif type had lower diversity than a region that participated equally in all motif types.

\begin{figure*}[t]
	\begin{center}
		\centerline{\includegraphics[width=1\textwidth]{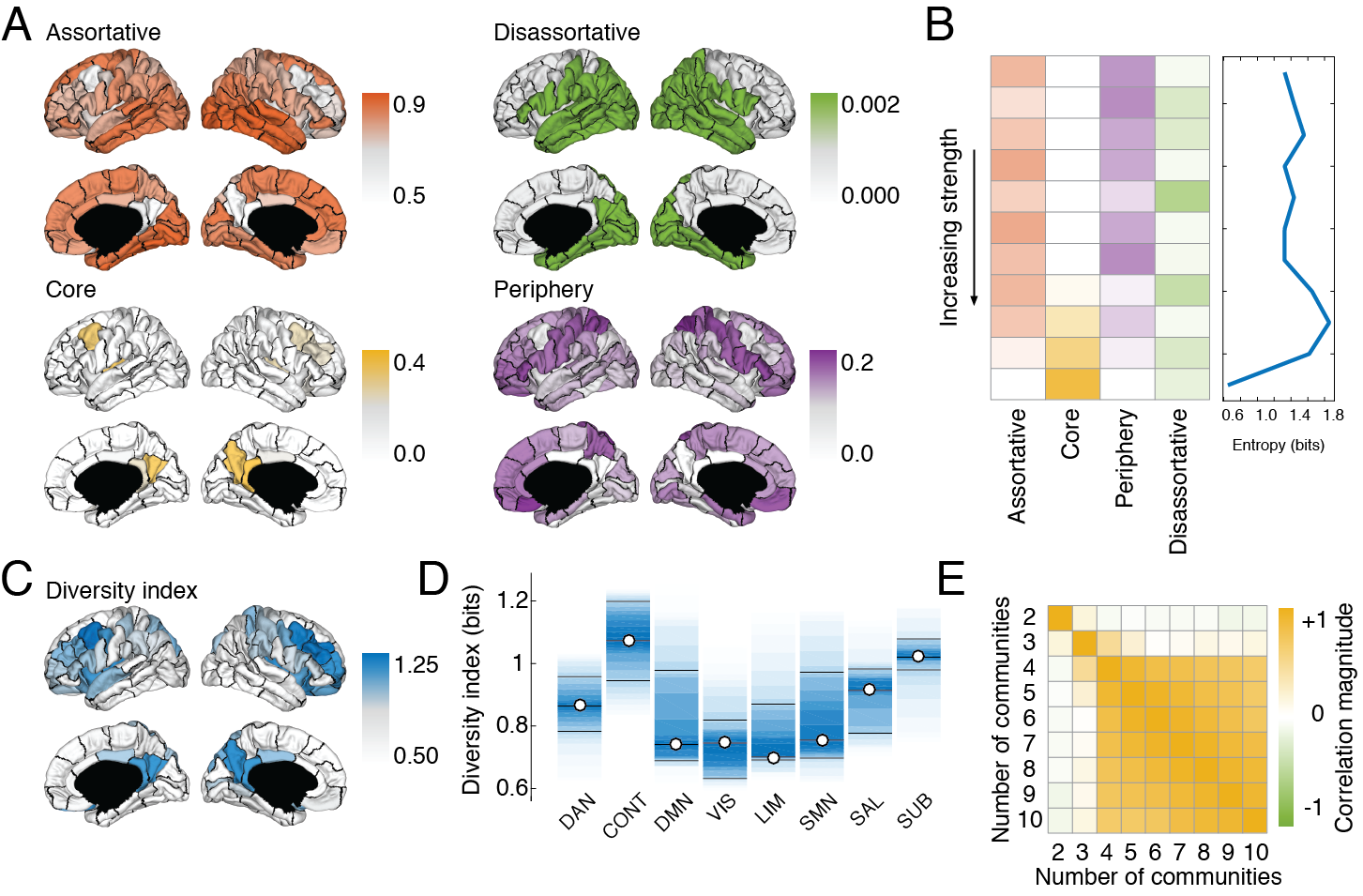}}
		\caption{\textbf{Regional variation in motif participation highlights diversely connected nodes.} (\emph{A}) Regional participation in the four community motif types. Note that the scales vary from panel to panel. (\emph{B}) Dominance of each motif type as a function of node strength (weighted degree). Note that motif dominance varies with strength; high-strength nodes are predominantly core while low-strength are assortative and peripheral. (\emph{C}) Diversity index measuring the entropy across each node's motif participation. (\emph{D}) Diversity index grouped by functional system. (\emph{E}) Correlation of regional diversity indices as a function of the number of detected communities, $K$, demonstrating robustness of results across choice of $K$.} \label{communityParticipation}
	\end{center}
\end{figure*}

As expected, motif participation was heterogeneous across brain regions. Core participation, for example, was dominated by highly connected sub-cortical regions as well as the precuneus, insula, and frontal cortices. This agrees with our understanding of those regions as being highly central in the network, with the capacity to exert influence and regulate information flow across the brain \cite{van2013network}. Participation in assortative motifs, on the other hand, was dominated by middle temporal, lateral occipital, and fusiform cortices (Fig.~\ref{communityParticipation}A). Interestingly, participation was stratified by node strength (Fig.~\ref{communityParticipation}B). Binning brain regions by their strengths, we found that low-strength bins were composed of nodes that participated predominantly in assortative and periphery motifs while high-strength bins included nodes that participated almost exclusively in core motifs. By quantifying the diversity within each bin as an entropy, we found that both the low- and high-strength regions were among the least diverse -- they participated in a narrow, well-defined set of motifs. The set of regions with the most diverse motif participation were those with above average but never the greatest strength (Fig.~\ref{communityParticipation}B). This finding was also observed in the non-human connectome datasets (Fig.~\ref{communityParticipationAllSpecies}).

These results suggest an important functional role for middle-strength brain regions. While both high- and low-strength regions are highly stereotyped in terms of the range of motifs in which they participate, middle-strength brain regions are among the most diverse, participating in all motif types nearly equally, and hinting at the capacity for enhanced functionality. This is not to diminish the putative functional roles of high- and low-strength regions, which have the capacity to readily exert influence and be influenced, respectively \cite{gu2015controllability}, but only to suggest that middle-strength nodes might have the ability to do both. Based on these findings, we hypothesized that the diversity of communities in which a region participates is related to its functional repertoire, with increased diversity corresponding to a broader range of functions. We further hypothesized that polymodal association areas, because they participate in a range of cognitive processes and require the synthesis of sensory information, attentional resources, and control mechanisms, would be among the most diverse.

To test this hypothesis, we computed a region-level diversity index (Fig.~\ref{communityParticipation}C), which we aggregated by functional system. As expected, the most diverse regions were concentrated within control and subcortical networks (permutation tests; $p = 0.001$) (Fig.~\ref{communityParticipation}D), and these results were robust across choice of the number of communities, $K$ (Fig.~\ref{communityParticipation}E). The cognitive control network includes some of the brain's more recently-evolved cortical structures \cite{buckner2013evolution} and are thought to play critical roles across a multitude of executive functions \cite{miller2000prefontral, botvinick2001conflict, dosenbach2007distinct, cole2013multi}, while the sub-cortex contains many nuclei responsible for performing distinct functional and regulatory roles \cite{squire1992memory, davis2001amygdala, grahn2008cognitive}. 


\subsection*{Behavioral relevance of motif diversity}

In the previous section, we demonstrated that the most diverse brain regions, in terms of their community motif participation, include control and subcortical regions. We speculated that this diversity might represent a neuroanatomical, network-level substrate for flexible cognitive behavior. In this section, we test whether inter-individual differences in regional diversity can account for behavioral variability.

Specifically, we asked 30 subjects to perform canonical cognitive control tasks (the Stroop and Navon), which require the rapid interactions of visual, attentional, and executive control systems (See \textbf{Materials and Methods} for task details). We combined total accuracy from both tasks to generate a composite accuracy score for each subject, which measured their performance. We hypothesized that individuals with greater diversity would perform better on both tasks than individuals with little diversity, suggesting that brain networks configured to facilitate integration across many types of meso-scale organization can more effectively exert control over processes requiring many complex representations and discriminative judgments.

To this end, we fit the WSBM to subjects' connectome data while varying the number of communities over the range $K = 2, \ldots, 10$, classified community motifs, and computed motif participation scores and diversity indices for each brain region. As an additional step, we partialed out the effect of subjects' total connection weight over the whole brain from the regional diversity indices. We then computed the Spearman correlation of total accuracy with the residuals, resulting in a correlation coefficient for each brain region.

\begin{figure*}[t]
	\begin{center}
		\centerline{\includegraphics[width=0.85\textwidth]{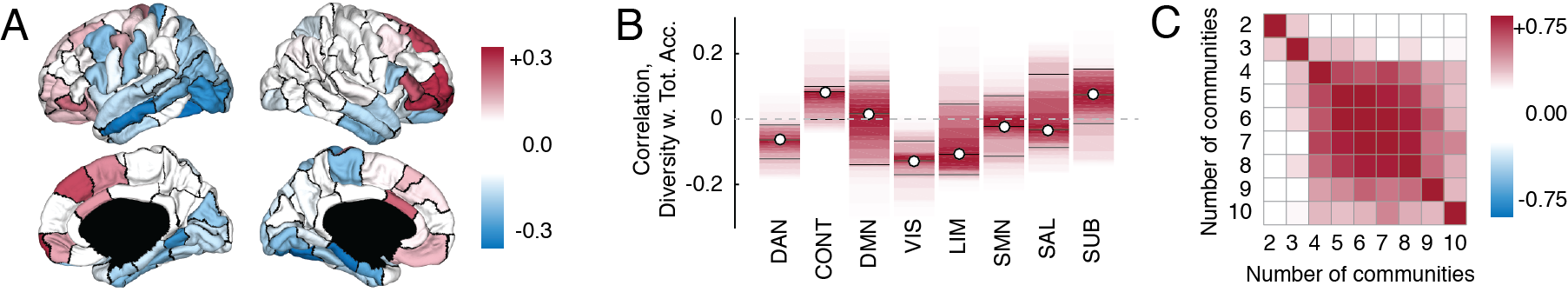}}
		\caption{\textbf{Diversity index correlates with individual differneces in performance on tasks demanding cognitive control.} (\emph{A}) Regional correlation coefficients of total accuracy with diversity index on the brain surface (shown with $K = 5$). Areas in white did not show a significant correlation after FDR correction for multiple comparisons. (\emph{B}) Regional correlation coefficients grouped according to functional system. DAN=dorsal attention, CONT=cognitive control, DMN=default mode, VIS=visual, LIM=limbic, SMN=somatomotor, SAL=salience, SUB=subcortical. (\emph{C}) Similarity of regional correlation coefficients as a function of the number of communities, indicating robustness to the choice of $K$ when $4 \leq K \leq 10$.} \label{behavioralAnalysis}
	\end{center}
\end{figure*}

Interestingly, we found that the strongest correlations (both positive and negative) were distributed heterogeneously across the brain (Fig.~\ref{behavioralAnalysis}A), but also tended to cluster within a few cognitive systems. The strongest positive correlations belonged to regions that were associated with cognitive control (permutation test; $p = 0.031$) and sub-cortical systems ($p = 0.015$), while the visual system was more anti-correlated than expected ($p < 10^{-3}$; all tests FDR corrected for multiple comparisons) (Fig.~\ref{behavioralAnalysis}B). Moreover, when the number of detected communities was greater than $K = 4$, this pattern remained largely unchanged (Fig.~\ref{behavioralAnalysis}C). Generally, these findings posit a link between individual differences in behavior and regional variation in motif diversity. More specifically, they also implicate control and subortical systems, a finding reminiscent of the cortico-striatal loops thought to play an important role in control-oriented behavior \cite{haber2016corticostriatal, medaglia2016functional}.


\section*{Discussion}

In this report, we hypothesized that the view of the connectome as being comprised of segregated communities is one that overlooks, for methodological convenience and ease of interpretation, competing and equally plausible accounts of its meso-scale architecture. These alternatives would allow for the possibility of more heterogeneous community configurations, including cores and peripheries and diassortative motifs. Using the weighted stochastic blockmodel, which belongs to a class of community detection algorithms sensitive to both assortative and non-assortative communities, we presented evidence supporting the existence of such motifs. Moreover, communities detected using the weighted stochastic blockmodel better recapitulated observed intrinsic functional relationships among brain regions in the human, and relationships between gene co-expression patterns in the mouse, compared to more commonly-used techniques such as modularity maximization. We then observed that the extent to which brain regions participated in particular classes of motifs varied across the brain, with the greatest degree of diversity concentrated in control and subcortical systems. This prompted us to formulate the hypothesis that participating in a diverse set of communities engenders a broader functional repertoire and flexible cognitive behavior. We tested this hypothesis using single-subject connectome data and behavioral scores recorded during cognitive control tasks. We found that regional diversity within both control and subcortical systems was predictive of subjects' task performance, supporting our hypothesis that structural diversity is relevant for human behavior.
 ~\\
~\\
\noindent \textbf{Topological support for cognitive processes} 
~\\
~\\
\noindent The relationship between form and function is one that has fascinated humanity throughout history. Within biology, this fascination motivates efforts to understand how cellular interaction pathways \cite{saadatpour2011dynamical,yeger2004network} or gene regulatory circuits \cite{adler2017optimal,shen2002network} impact a host of biological processes. In neural systems in particular, evidence across spatial scales of measurement point to the critical importance of structural links between neural elements in predicting the function of a circuit or motif \cite{sporns2004motifs, marrelec2016functional}. For example, the structural dynamics of synapse formation can split functional cell-assemblies into subassemblies, thereby producing distributed synchrony \cite{levy2001distributed}. Indeed, the structural constraints of synapse existence and formation offer a mechanistic explanation for sequential neural activity patterns necessary for motor gestures and sequential cognitive processes such as memory \cite{rajan2016recurrent}; evidence from theoretical and computational studies suggests that neural sequences may be shaped by synaptic constraints and network circuitry rather than cellular time constants \cite{fiete2010spike}. At the small-scale of neuronal circuits, these studies collectively higlight topological rules leading to structural motifs that explain observed function: from feedforward loops \cite{gollo2014frustrated} to repressor lattices \cite{jensen2009repressor}. Yet obtaining similar insights at the large-scale of whole-brain circuits has proven challenging \cite{hermundstad2013structural,hermundstad2014structurally}, in part due to the lack of an understanding regarding what motifs matter at this scale \cite{sizemore2016cliques}. Our work directly addresses this gap by offering novel concepts to define structural motifs in the meso-scale architecture of connectomes, tools for estimating these motifs computationally from available data, and a proof-of-concept demonstration that the diversity of such motifs can be used to understand individual differences in a cognitive process that requires computations in large-scale distributed circuits (cognitive control \cite{botvinick2014computational}). 
~\\
~\\
\textbf{Community and meso-scale connectome analyses}
~\\
~\\
\noindent In seeking computationally-relevant large-scale structural motifs, we shifted focus away from regional and whole-brain organization, and onto its meso-scale structure \cite{betzel2016multi}. Meso-scale network analysis is a coarse-graining of network features analogous to dimension reduction. While the modest-sized networks studied here benefit from such an analysis, this approach may see broader applicability in the future, where advances in connectome imaging and reconstruction techniques have resulted in high-dimensional, high complexity datasets \cite{zador2012sequencing, kasthuri2015saturated, kim2016long}. Making sense of such data while still respecting its underlying relational structure -- which can be parsimoneiously encoded in a network or graph representation -- is a major computational challenge. To comprehend the organization of connectome data, especially at the cellular scale, requires dimension reduction techniques like community detection that can distill the important organizational principles from those that are less useful. Modularity maximization and related techniques, while widely-used now, may miss important and functionally-relevant characteristics of the complicated architecture of neural circuits, which exhibit non-assortative, cell type-specific wiring diagrams \cite{real2017neural, briggman2011wiring, ding2016species}. Our study represents one of the first to explore the utility of blockmodels in conjunction with human and animal data \cite{simpson2011exponential, simpson2012exponential, pavlovic2014stochastic, moyer122015mixed, murphy2016explicitly, baldassano2015parcellating}, providing the opportunity to explicitly test its utility in explaining functional dynamics (in the human) and gene co-expression (in the mouse). Future studies may wish to extend these approaches to the study of neurodevelopment \cite{martino2014unraveling}, or the alteration of connectomic structure in psychiatric disease \cite{rubinov2013fledgling,fornito2012schizophrenia} and neurological disorders \cite{bassett2009human,stam2014modern}.
~\\
~\\
\textbf{Connectomes exhibit rich, non-assortative structure}
~\\
~\\
\noindent In revealing non-assortative meso-scale structure in human and animal connectomes, our work refines the widely-held view of the brain as being composed of segregated communities interlinked by integrative hubs \cite{van2013network}. While most commmunities detected using the weighted stochastic blockmodel engaged in assortative interactions (and were hence segregated from one another), many also engaged in core-periphery interactions and a few engaged in disassortative interactions. This diversity of interconnection types suggests that the integrative units of the brain are not necessarily highly-connected hubs, but groups of brain regions with similar connectivity profiles and weight distributions with shared functional capacity \cite{warren2014network}. Importantly, these findings were replicated in non-human connectome datasets including \emph{Drosophila}, mouse, rat, and macaque, suggesting that non-assortative meso-scale architecture is not unique to \emph{Homo sapiens} and is, likely, not an artifact attributable to a specific connectome reconstruction technique. The richness of assortative, disassortative, and core-periphery mesoscale interactions inherently alters the picture of possible cognitive computations that the system can support. Indeed, that picture morphs from a simple egalitarian description in which each cognitive system acts equally and independently \cite{fodor1983modularity}, into a more varied landscape that supports top-down \cite{dalley2011impulsivity} and bottom-up influence \cite{fries2015rhythms}, hierarchies of processes \cite{chaudhuri2014diversity,felleman1991distributed}, and repertoire diversity \cite{deco2012ongoing}, all canonical features of neural dynamics observed in empirical studies across spatial scales and species.
~\\
~\\
\textbf{A critical role for the topological middle class}
~\\
~\\
\noindent By using a computational approach sensitive to meso-scale diversity (weighted stochastic block models), and developing computational methods to exercise that approach in human and animal connectomes (community morphospace), we were able to uncover an unexpectedly critical role of the topological middle class -- brain regions with above average but never the greatest strength displayed the most diverse motif participation. This finding joins a series of recent examples highlighting the importance of brain regions that are not the high-strength hubs or premier members of the so-called \emph{rich-club}, as well as those explaining globally correlated brain states by local, weak pairwise interactions \cite{schneidman2006weak}. Indeed, several recent functional studies have demonstrated the unexpected utility of low-strength nodes in explaining individual differences in fluid intelligence \cite{cole2012global,santarnecchi2014efficiency} and alterations in functional brain dynamics in psychiatric disease \cite{bassett2012altered}. Complementary studies of structural architecture in the human and non-human primate connectomes suggest that low-strength nodes have the capacity to drive neural dynamics into particularly distant target states \cite{gu2015controllability,muldoon2016stimulation,tang2016developmental}. Yet, the role of middle-strength nodes has not been broadly studied, and the architectural features characteristic of their structure are not well understood. Our finding that middle strength nodes display diverse participation in all motif types suggests the capacity for enhanced functionality: the capacity to both readily exert influence on and be influenced by others \cite{gu2015controllability}. While we provide an empirical observation of this diversity, and initial evidence supporting its role in cognitive function (see next section), our understanding could benefit from future work exploring mathematical models of neural dynamics that explicitly account for these observations.
~\\
~\\
\textbf{Community diversity predicts cognitive performance}
~\\
~\\
We demonstrated that individual differences among brain regions' diversity was predictive of subjects' performances on Stroop and Navon tasks. Moreover, the most predictive regions were associated with the systems whose diversity was, on average, the greatest (specifically, the control and sub-cortical networks). This finding suggests that diverse motif participation, which we speculate allows regions to engage in a wider range of function, represents a neuroanatomical substrate for flexible cognitive behavior \cite{cole2013multi, mohr2016integration}. This finding corroborates past studies showing that the brain's control systems reorganize the flexible reconfiguration of brain FC in adapting to complex tasks \cite{cummings1993frontal, parent1995functional, forstmann2010cortico}.

The fact that sub-cortical and control systems are, on average, most predictive of cognitive performance agrees closely with what is already known about their circuitry; cortico-subcortical loops linking them to one another are believed to play important roles in supporting cognitive control processes \cite{heyder2004cortico, wei2016inhibitory}. In our work, these systems are also highlighted as the most diverse in terms of their motif participation, which suggests that in addition to their traditional roles in cognitive control, they may also have the capacity to perform manifold functional roles, including the support of network-wide polysensory integration and association as well as exerting control over those processes. Future studies can explicitly examine whether this meso-scale anatomical organization guides functional network organization such as those revealing a heirarchy from primary sensory to associative multimodal to executive control regions \cite{sepulcre2012stepwise}.

~\\
~\\
\textbf{Broad utility of novel network-based metrics}
~\\
~\\
\noindent With the rapid acquisition of connectomic data across species and spatial scales \cite{bassett2017network}, as well as other relational data in genetics \cite{norton2016detecting}, molecular biology \cite{conaco2012functionalization}, materials \cite{zhang2016stretch}, and social systems \cite{schmaelzle2017brain}, it is becoming increasingly urgent to develop and share computational tool sets to distill organizational principles in complex systems \cite{glaser2016development,marblestone2016toward}. Over the course of this report we made several methodological innovations in the form of novel network-based metrics, including assortativity scores, community morphospace analysis and motif classification, and node-level motif participation and diversity indices. With the exception of the diversity index, our exploration of these measures was largely theoretical. Future studies can capitalize on these theoretical advances by comparing their values in healthy and diseased conditions \cite{fornito2015connectomics}, across development, senescence, and the human lifespan \cite{zuo2016human}, or as a function of cognitive capacity or behavioral state \cite{barch2013function}. Moreover, the tools themselves are agnostic to the exact nature of the network under study, and may therefore also prove useful in understanding meso-scale organization in physical \cite{bassett2015extraction,papadopoulos2016evolution} and technological \cite{onnela2016harnessing} systems.

\subsection*{Methodological Considerations and Limitations}

It is important to note a few methodological limitations of this study. First, the WSBM requires that the user specify the number of communities, $K$. While the optimal $K$ can be estimated using any number of heuristic approaches, how to do so in a principled statistical manner is regarded as an open problem \cite{jacobs2014unified, peel2016ground}. Accordingly, we opted not to focus on any particular value of $K$, but to show that our results are robust over a reasonable choice of $K$. Another limitation, especially with the human connectome data, is the versimilitude of the reconstructed network. Diffusion imaging and tractography algorithms are prone to inaccurate reconstructions that limit their utility for connectome inference \cite{thomas2014anatomical, reveley2015superficial, maier2016tractography}. Despite these shortcomings, tractographic reconstructions of the brain's white matter pathways have been incorporated into neurosurgical planning \cite{yu2005diffusion,fernandez2012high}, suggesting that in specific contexts (i.e. for particular white-matter pathways), tractography represents an accurate method for \emph{in vivo} network mapping. Moreover, hardware advances and a new generation of ensemble \cite{takemura2016ensemble} and global reconstruction techniques \cite{pestilli2014evaluation, smith2013sift} offer the possibility of improved estimates. In the context of this discussion, however, it is also important to point out that our results in the human were all confirmed in the non-human connectomes as well, which are constructed from inherently different sorts of empirical data. The reliability of our findings across \emph{Drosophila}, mouse, rat, macaque, and human suggest that they cannot be accounted for by deficiencies in any one data modality. 

\subsection*{Conclusion}

In this work, we sought to understand the structural basis for cognitive computations. We hypothesized that diverse meso-scale structure allows a network to engage in a wider functional repertoire, and that inter-subject variability in diversity is predictive of variation in cognitive performance. To address these hypotheses, we applied a weighted stochastic blockmodel to the connectome data acquired from five different species (\emph{Drosophila}, mouse, rat, macaque, and human). We showed that the communities it detects are different from those commonly discussed in the literature, and that they provide statistically better explanations of resting state functional connectivity in the human and gene co-expression in the mouse. Finally, we showed that a diversity metric derived from those communities predicts behavioral outcomes in cognitive control tasks. Collectively, this body of work provides an aternative view of the structural substrate for computations in large-scale distributed circuits, and opens up new avenues of inquiry into the development and evolution of this architecture.

\section*{Materials and Methods} \label{Materials and Methods}

\subsection*{Connectome datasets}

A \emph{connectome} refers to the complete set of neural elements and the physical connections that link those elements to one another \cite{sporns2004organization}. We analyzed previously-published connectome data representative of five different species: \emph{Drosophila}, mouse, rat, macaque, and human. In this section we offer brief descriptions of the methodologies used to reconstruct each connectome alongside their original reference(s).

\subsubsection*{Human}
We analyzed both individual and group-representative, whole-brain networks generated by combining single-subject data from a cohort of 30 healthy adult participants. Each participant's network was reconstructed from diffusion spectrum images (DSI) in conjunction with state-of-the-art tractography algorithms to estimate the location and strength of large-scale interregional white-matter pathways. Study procedures were approved by the Institutional Review Board of the University of Pennsylvania, and all participants provided informed consent in writing. Details of the acquisition and reconstruction have been described elsewhere \cite{betzel2016optimally, betzel2016modular}. We studied a division of the brain into $N=128$ brain regions (nodes) \cite{cammoun2012mapping}. Based on this division, we constructed for each individual an undirected and weighted connectivity matrix, $A \in \mathbb{R}^{N \times N}$, whose edge weights were equal to the number of streamlines detected between regions $i$ and $j$ normalized by the geometric mean of their volumes: $A_{ij} = \frac{S_{ij}}{\sqrt{(V_i V_j)}}$.

\noindent The resulting network was undirected (i.e. $A_{ij} = A_{ji}$). These individual-level networks were then aggregated to form a group-representative network. This procedure can be viewed as a distance-dependent consistency thresholding of connectome data and the details have been described elsewhere \cite{mivsic2015cooperative, betzel2016modular}. The resulting group-representative network has the same number of binary connections as the average individual and the same edge length distribution. This type of non-uniform consistency thresholding has been shown to be superior to other, more commonly used forms \cite{roberts2016consistency}.

\subsection*{Behavioral tasks}
All participants completed a modified local-global perception task based on classical Navon figures \cite{navon1977forest} and a Stroop task with color-word pairings that were eligible and ineligible to elicit interference effects \cite{van2005separating}. 

For the Navon task, local-global stimuli were comprised of four shapes -- a circle, X, triangle, or square -- that were used to build the global and local aspects of the stimuli. On all trials, the local feature did not match the global feature, ensuring that subjects could not use information about one scale to infer information about another. Stimuli were presented on a black background in a block design with three blocks. In the first block type, subjects viewed white local-global stimuli. In the second block type, subjects viewed green local-global stimuli. In the third block type, stimuli switched between white and green across trials uniformly at random with the constraint that 70\% of trials included a switch in each block. In all blocks, subjects were instructed to report only the local features of the stimuli if the stimulus was white and to report only the global feature of the stimuli if the stimulus was green. Blocks were administered in a random order. Subjects responded using their right hand with a four-button box. All subjects were trained on the task outside the scanner until proficient at reporting responses using a fixed mapping between the shape and button presses (i.e., index finger = ``circle'', middle finger = ``X'', ring finger = ``triangle", pinky finger = ``square''). In the scanner, blocks were administered with 20 trials apiece separated by 20 s fixation periods with a white crosshair at the center of the screen. Each trial was presented for a fixed duration of 1900 ms separated by an interstimulus interval of 100 ms during which a black screen was presented.

For the Stroop task, trials were comprised of words presented one at a time at the center of the screen printed in one of four colors -- red, green, yellow, or blue -- on a gray background. For all trials, subjects responded using their right hand with a four-button box. All subjects were trained on the task outside the scanner until proficient at reporting responses using a fixed mapping between the color and button presses (i.e., index finger = ``red'', middle finger = ``green'', ring finger = ``yellow", pinky finger = ``blue'').  Trials were presented in randomly intermixed blocks containing trials that were either eligible or ineligible to produce color-word interference effects. In the scanner, blocks were administered with 20 trials apiece separated by 20 s fixation periods with a black crosshair at the center of the screen. Each trial was presented for a fixed duration of 1900 ms separated by an interstimulus interval of 100 ms during which a gray screen was presented. In the trials ineligible for interference, the words were selected to not conflict with printed colors (``far,'' ``horse,'' ``deal,'' and ``plenty''). In the trials eligible for interference (i.e. those designed to elicit the classic Stroop effect \cite{stroop1935studies}), the words were selected to introduce conflict (i.e. printed words were ``red,'' ``green,'' ``yellow,'' and ``blue'' and always printed in an incongruent color). 

\subsubsection*{Additional data}
The connectome data was accompanied by (1) annotated system labels, which assigned each node to a single functional system, and (2) a group-representative functional connectivity (FC)  matrix constructed from resting state scans which were collected concurrently with the behavioral data. See \cite{medaglia2016functional} for details. The system labels were taken from \cite{yeo2011organization} and included seven cortical systems (dorsal attention, control, default mode, visual, limbic, somatomotor, and salience networks) along with an eighth sub-cortical label. The group-representative resting state FC network was generated by averaging subject-level resting state FC and by partialling out the effect of distance. The elements of the resulting matrix quantified the strength of functional connection between brain regions beyond what would be expected given their Euclidean distance from one another.
 
\subsection*{Stochastic blockmodel}
The stochastic blockmodel (SBM) seeks to partition a network's nodes into $K$ communities. Let $z_i \in \{1 , \ldots, K\}$ indicate the community label of node $i$. Under the standard blockmodel, the probability that any two nodes, $i$ and $j$, are connected to one another depends only on their community labels: $p_{ij} = \theta_{z_i,z_j}$.
 
 To fit the blockmodel to observed data, one needs to estimate the parameters $\theta_{rs}$ for all pairs of communities $\{r,s\} \in \{1 , \ldots, K\}$ and the community labels $z_i$. Assuming that the placement of edges are independent of one another, the likelihood of a blockmodel having generated a network, $A$, can be written as:
 
 \begin{equation}
 P(A | \{ \theta_{rs} \}, \{ z_i \} ) = \prod_{i,j > i}  \theta_{z_iz_j}^{A_{ij}}(1-\theta_{z_iz_j})^{1-A_{ij}} .
 \label{eq:1}
 \end{equation}
 
 \noindent Fitting the SBM to an observed network involves selecting the parameters $\{ \theta_{rs} \}$ and $\{ z_i \}$ so as to maximize this function.
 
 \subsection*{Weighted stochastic blockmodel}
 The classical SBM is most often applied to binary networks where edges carry no weights. In order to maximize their utility to the network neuroscience community (where most networks are weighted), the SBM needs to be able to efficiently deal with weighted edges. Recently, the binary stochastic blockmodel was extended to weighted networks as the weighted stochastic blockmodel (WSBM) \cite{aicher2013adapting, aicher2014learning}. 
 
 Equation \ref{eq:1} can be rewritten in the form of an exponential family of distributions \cite{aicher2013adapting}:
 
 \begin{equation}
 P(A | \{ \theta_{rs} \}, \{ z_i \} ) \propto \exp \Bigg( \sum_{ij} T(A_{ij}) \cdot \eta(\theta_{z_iz_j}) \Bigg) .
 \end{equation}
 
 \noindent For the classical (unweighted) SBM, $T$ is the sufficient statistic of the Bernoulli distribution and $\eta$ is its function of natural parameters. Different choices of $T$ and $\eta$, however, can allow edges and their weights to be drawn from other distributions. The WSBM, like the classical SBM, is parameterized by the set of community assignments, $\{ z_i \}$, and the parameters $\theta_{z_i z_j}$. The only difference is that $\theta_{z_i z_j}$ now specifies the parameters governing the weight distribtion of the edge, $z_i z_j$.
 
 Here, we follow \cite{aicher2013adapting}, and model edge weights under a normal distribution, whose sufficient statistics are $T = (x,x^2,1)$ and natural parameters $\eta = (\eta/\sigma^2, -1/(2\sigma^2),-\mu^2/(2\sigma^2))$. Under this distribution, the edge $z_i z_j$ is parameterized by its mean and variance, $\theta_{z_i z_j} = (\mu_{z_i z_j}, \sigma_{z_i z_j}^2)$, and the likelihood is given by:
 
 \begin{equation}
 P(A | \{ z_i \}, \{ \mu_{rs} \}, \{ \sigma_{rs}^2 \} ) = \prod_{ij} \exp \Bigg(  A_{ij} \cdot \frac{\mu_{z_iz_j}}{\sigma_{z_iz_j}^2} - A_{ij}^2 \cdot \frac{1}{2 \sigma_{z_i z_j}^2} - 1 \cdot \frac{\mu_{z_i z_j}^2}{\sigma_{z_i z_j}^2}\Bigg).
 \end{equation}
 
 The above form assumes that all possible edges falling between communities are drawn from a normal distribution. However, most connectomes are sparse, i.e. edges where $A_{ij} = 0$ indicate the absence of a connection. One solution for dealing with this problem is to model edge weights with an exponential family distribution and to model the presence or absence of edges by a Bernoulli distribution (akin to the unweighted SBM) \cite{aicher2013adapting}. Letting $T_e$ and $\eta_e$ represent the edge-existence distribution and $T_w$ and $\eta_w$ represent the normal distribution governing edge weights, we can rewrite the likelihood function for the sparse WSBM as:
 
 \begin{equation}
 \log( P(A | z , \theta ) ) = \alpha \sum_{ij \in E} T_e(A_{ij}) \cdot \eta_e (\theta_{z_i z_j}^(e)) + (1 - \alpha) \sum_{ij} T_w (A_{ij}) \cdot \eta_w ( \theta_{z_i z_j}^{w} ) ~,
 \end{equation}
 
 \noindent where $E$ is the set of all possible edges, $W$ is the set of weighted edges ($W \subset E$), and $\alpha \in [0,1]$ is a tuning parameter governing the relative importance of either edge weight or edge presence (or absence) for inference. Here, we fix $\alpha = 0.5$, which balances their relative importance.
 
 For each of the five datasets (connectomes from \emph{Drosophila}, mouse, rat, macaque, and human), we maximize the likelihood of this sparse WSBM using a Variational Bayes technique described in \cite{aicher2013adapting} and implemented in MATLAB using code made available at the author's personal website (\url{http://tuvalu.santafe.edu/~aaronc/wsbm/}). We varied the number of communities from $K = 2, \ldots , 10$ and repeated the optimization procedure 250 times, each time initializing the algorithm with a different set of parameters.
 
 \subsection*{Modularity maximization}
 Blockmodels are flexible and can accommodate various classes of community structure. In network neuroscience \cite{bassett2017network}, however, the majority of studies examining the brain's community structure have focused on its division into assortative communities by maximizing a modularity quality function:
 
 \begin{equation}
 Q(\{ z_i \}, \gamma) = \sum_{ij} [A_{ij} - \gamma \cdot P_{ij}] \delta(z_i z_j).
 \end{equation}
 
\noindent Here, $P_{ij}$ is the expected number of connections between nodes $i$ and $j$ under a null connectivity model and $\delta(\cdot \cdot)$ is the Kronecker delta function and is equal to $1$ when its arguments are the same and $0$ otherwise. $Q(\{ z_i \}, \gamma)$ is maximized by choosing community assignments $z_i$ that result in modules whose observed internal density maximally exceeds what would be expected under the null model. The free parameter, $\gamma$, is the structural resolution parameter and can be tuned to uncover communities of different size \cite{reichardt2006statistical, fortunato2007resolution}. The partition $\mathcal{P} = \{ z_i \}$ that maximizes $Q(\{ z_i \}, \gamma)$ is usually treated as a reasonable estimate of the network's community structure. While recent studies have investigated alternative definitions of $P_{ij}$ \cite{bassett2013robust,bassett2015extraction,papadopoulos2016evolution}, we use the common configuration model: $P_{ij} = \frac{k_i k_j}{2m}$, where $k_i = \sum_j A_{ij}$ and $2m=\sum_i k_i$ \cite{betzel2016modular}.
 
 Unlike WSBMs, most modularity maximization algorithms (henceforth referred to as $Q_{max}$) do not allow the user to specify the number of detected communities. In order to extract partitions of the network into exactly $K$ communities, we proposed a greedy algorithm in which nodes are initialized with random $K$-community partition, their assignments switched one at a time, and the new assignment accepted if the switch results in an increased $Q$. We repeated this algorithm 250 times for each $K$ and during each repetition considered 10000 random community switches. We fixed $\gamma = 1$ throughout.

\subsection*{Statistics for comparing the WSBM with $Q_{max}$}
\subsubsection*{Variation of information}
Modularity maximization is designed to uncover assortative communities while blockmodels are capable, at least in principle, of detecting more general types of community structure. It is unclear, however, when applied to brain network data whether the detected communities using either technique will actually differ from one another. We develop a set of statistics for comparing community structure at different topological scales ranging from global (whole partition), to mesoscale (community), to local (individual node).

At the global scale, we compare two partitions, $\mathcal{P}_1 = \{ z_i^1 \}$ and $\mathcal{P}_2  = \{ z_i^2 \}$, using the dissimilarity measure \emph{variation of information}, $VI$, which yields an information theoretic distance between two partitions \cite{meilua2003comparing}:

\begin{equation}
VI(\mathcal{P}_1,\mathcal{P}_2) = H(\mathcal{P}_1) + H(\mathcal{P}_2) - 2I(\mathcal{P}_1,\mathcal{P}_2)~\\,
\end{equation}

\noindent where $H(\mathcal{P})$ and $I(\mathcal{P},\mathcal{Q})$ are the entropy and mutual information. The more similar two partitions are to one another, the closer their variation of information is to zero. Two partitions may differ from one another, trivially, if they feature a different number of communities. Throughout this section and the next and in order to avoid this issue, we only compare partitions if they feature the same total number of communities.

\subsubsection*{Community and regional assortativity}

While variation of information makes it possible to assess the similarity of partitions as a whole, we also wanted to assess which brain regions, systems, and communities differ between techniques. One dimension along which we expect the techniques to differ is the extent to which the detected communities are assortative. To quantify this property, we propose community and regional \emph{assortativity scores}.

For a community $r$, we define its assortativity as:

\begin{equation}
\mathcal{A}_r = [\omega_{rr} - \max_{s \ne r} (\omega_{rs})],
\end{equation}

\noindent where $\omega_{rs} = \frac{1}{n_r \cdot n_s} \sum_{i \in r, j \in s} A_{ij}$ is the weighted density of connections between communities $r$ and $s$. For directed networks, we consider both incoming and outgoing connections, and we replace $\max_{s \ne r} (\omega_{rs})$ with the greater of $\max_{s \ne r} (\omega_{rs}^\text{In})$ or $\max_{s \ne r} (\omega_{rs}^\text{Out})$.

We also calculated an analogous score for individual brain regions. Given region $i$'s community assignment $z_i$, we calculated its connection density to community $r$ as $a_{ir} = \frac{1}{n_r} \sum_{j \in r} A_{ij}$. Then, its regional assortativity score was given by:

\begin{equation}
a_i = a_{i z_i} - \max_{r \ne z_i} a_{ir}~\\.
\end{equation}

\noindent Again, we modified this equation slightly for directed networks to take into account both incoming and outgoing connections. We replaced $a_{i z_i}$ with the lower of either $a_{iz_i}^\text{In}$ or $a_{i z_i}^\text{Out}$, and we replaced $\max_{r \ne z_i} a_{ir}$ with the greater of either $\max_{r \ne z_i} a_{ir}^\text{In}$ or $\max_{r \ne z_i} a_{ir}^\text{Out}$.

Under this definition, the assortativity score measures the minimum difference between the density of connections made between a region and its own community, and the density of connections made between a region and any other community. In computing both regional and community assortativity scores, we excluded singleton communities.

\subsection*{Maximally assortative set}
In addition to the metrics described above, we also sought to identify the largest set of communities uncovered by the WSBM that exhibits assortative community structure. We termed this set the \emph{maximally assortative set} and defined it as the set of $k \le K$ communities, $\{ c_1 , \ldots , c_k \}$ such that $\displaystyle \min_i (\omega_{c_i,c_i}) > \max_{i\ne j} (\omega_{c_i,c_j})$ and the total number of nodes in those communities was maximized.

\subsection*{Community interaction motifs and morphospace analysis}
Uncovering a network's community structure makes it possible to shift focus away from individual nodes and edges and onto communities and their aggreate interactions with one another. Taking such a coarse view of a network can make it possible to more easily infer the functions of communities and the roles of individual nodes within those communities.

Here, we study those interactions using a \emph{theoretical morphospace analysis} \cite{mcghee2006geometry}, a technique recently adapted to the study of complex networks \cite{avena2014using,avena2015network,goni2013exploring}. A morphospace is a hyperspace whose axes represent the features of an organism or system. Take, for example, \emph{foraminiferal} tests -- the shells that form the outer layers of certain aquatic protists -- that can be modeled and fully parameterized using a small number of morphological traits \cite{tyszka2006morphospace}. A simple morphospace can be constructed whose axes are represented by these traits, and any observed test can then be situated within this space. Oftentimes, there will exist certain regions of space (i.e. particular sets of traits) that are densely populated and other regions that, by comparison, are not populated at all. By studying which sets of traits are more common, it becomes possible to deduce the evolutionary constraints and pressures that drove their emergence.

It is in this same spirit that network morphospaces can be constructed \cite{avena2015network}. Instead of axes representing an organism's morphological or physiological traits, the axes of a network morphospace represent topological properties of a network, e.g. its efficiency, wiring cost, complexity, etc. \cite{goni2013exploring}, or the parameters of network models \cite{vertes2012simple, betzel2016generative}. 

In this case, we construct a \emph{community morphospace}. Each point in the morphospace represents a pair of communities, $r$ and $s$, and the point's location is given by the within- and between-community connection densities: $\omega_{rr}$, $\omega_{ss}$, and $\omega_{rs}$. Given these values, we can also classify community interactions into one of three distinct motifs (interaction types):

\[
M_{rs} = 
\begin{cases}
M_\text{assortative},& \text{if } \min( \omega_{rr},\omega_{ss} ) > \omega_{rs}\\
M_\text{core-periphery},& \text{if } \omega_{rr} > \omega_{rs} > \omega_{ss}\\
M_\text{core-periphery},& \text{if } \omega_{ss} > \omega_{rs} > \omega_{rr}\\
M_\text{disassortative},& \text{if } \omega_{rs} > \max( \omega_{rr},\omega_{ss} ).
\end{cases}
\]

\noindent From these classifications we were able to associate motifs to individual nodes. Node $i$'s participation in motif $M$ was calculated as the number of times that the community to which node $i$ was assigned interacted with any other community to form a motif of type $M$. We then normalized these counts by the total number of motifs (for a $K$-community partition there are in total $K(K - 1)/2$ or $K(K - 1)$ total motifs depending upon whether the network is undirected or directed, respectively). Importantly, when computing participation in core-periphery motifs, we distinguished between the core and periphery, and computed separate participation scores for each. Finally, from participation types we computed each node's \emph{diversity index}, which measured the entropy of its normalized participation in each motif type.

\section*{Acknowledgements}
We are grateful to Mikail Rubinov for sharing mouse connectome and gene expression data, and to Olaf Sporns for sharing rat connectome data.

\newpage
\bibliography{../sbm_connectome_biblio}

\newpage
\beginsupplement

\section*{Supplementary materials}

In this supplement, we describe four additional non-human connectome datasets which we analyzed using the same methods and techniques as the human data. Whereas the human connectome data described in the main text were reconstructed \emph{in vivo} from diffusion weighted images using deterministic tractography algorithms, the procedures used to reconstruct the non-human data were more varied, ranging from retrograde tract tracing to meta-analysis. These data allow us to assess whether the differences we observed between the WSBM and $Q_{max}$ techniques are a consequence of the procedures used to reconstruct the connectome data. In general, the analyses we present in this supplement corroborate those in the main text.

\subsection*{Non-human connectome data}

\begin{figure*}[t]
	\begin{center}
		\centerline{\includegraphics[width=1\textwidth]{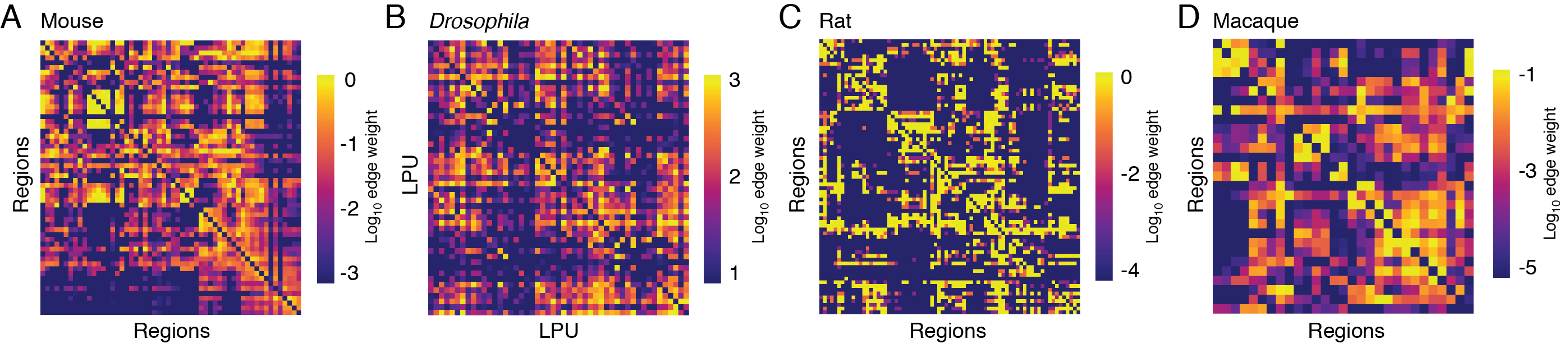}}
		\caption{\textbf{Non-human connectome data.} Connectivity matrices for the four non-human connectome datasets.} \label{basicNetworkComparisonNonHuman1}
	\end{center}
\end{figure*}

\subsubsection*{Mouse}
In addition to human data, we analyzed a mouse connectome reconstructed from tract-tracing experiments made publicly available by the Allen Brain Institute \cite{oh2014mesoscale}. Tracers were tracked from a series of injection sites to ispi- and contra-lateral brain regions. We focus on a reprocessed version of these data in which the mouse brain was parcellated into $N = 112$ regions (56 per hemisphere) and edge weights defined as volume-normalized number of connections between regions \cite{rubinov2015wiring}. Rather than use the whole-brain connectome described in that report (generated by copying and mirroring tract-tracing data from one hemisphere to the other), we analyzed a single cerebral hemisphere ($N = 56$ regions) for which complete connectivity data was availabe. Due the directed nature of the tract-tracing experiments, the resulting network was asymmetric (i.e. $A_{ij} \ne A_{ji}$) (Fig.~\ref{basicNetworkComparisonNonHuman1}A).

\subsubsection*{Drosophila}
We also analyzed a network reconstructed from 12,995 projection neurons in the female \emph{Drosophila} brain \cite{chiang2011three, shih2015connectomics}. Neurons were aggregated among $N = 49$ local processing units (LPUs), which represent network nodes. The resulting network is directed and weighted and has been reported elsewhere \cite{shih2015connectomics,worrell2017signal} (Fig.~\ref{basicNetworkComparisonNonHuman1}B).

\subsubsection*{Rat}
We also analyzed a rat cortical network \cite{bota2015architecture}. This network was constructed by collating reports on rat tract-tracing experiments \cite{bota2005brain}, extracting information from those reports regarding the existence of connections (resulting in $>$16000 connections), and based on the consistency and quality of those results, assigning a single weight to existing inter-regional connections. The result is a directed network of $N = 73$ cortical regions spanning a single hemisphere (Fig.~\ref{basicNetworkComparisonNonHuman1}C).

\subsubsection*{Macaque}
We also analyzed an incomplete macaque connectome documenting the connections among $N = 29$ cortical regions \cite{markov2012weighted,markov2013role}. As with some of the other datasets, the macaque connections were established from retrograde tract-tracing experiments and were weighted as fraction of labeled neurons (``the number of labeled neurons in a given source area relative to the total number of labeled neurons in the brain for any given injection''). The result is a weighted and directed connectivity matrix (Fig.~\ref{basicNetworkComparisonNonHuman1}D). 

\subsection*{Additional data}
As with the human dataset, each non-human dataset was accompanied by system labels. For the mouse data, we used the high-resolution parcellation of the mouse brain into eight systems described in \cite{rubinov2015wiring} (three subdivisions of an olfactory/hippocampal network, a somatomotor/sensory/motor system, a brainstem/cerebellar system, and three subdivisions of an audition/visual network). For \emph{Drosophila} we used a division of the network into five systems (audition/mechanosensation, olfactory, premotor, visual-left, visual-right) \cite{shih2015connectomics}. For rat we used a division into three systems (sensory/motor, poly-association, and cortical subplate) \cite{bota2015architecture}. For macaque we used a division into six systems \cite{markov2013cortical} (prefrontal, pariteal, temporal, occipital, frontal, and other).

Finally, the mouse dataset included gene expression profiles for each region. These data were originally made available by the Allen Brain Institute and were obtained using high-throughput \emph{in situ} hybridization and acquization, and included 3,380 of the $>$ 20,000 genes assayed that survived quality control. See \cite{rubinov2015wiring} for criteria for inclusion of this particular set of genes. From these gene expression data, we computed the correlation of each pair of regions' expression profiles, resulting in a square matrix.

\begin{figure*}[t]
	\begin{center}
		\centerline{\includegraphics[width=1\textwidth]{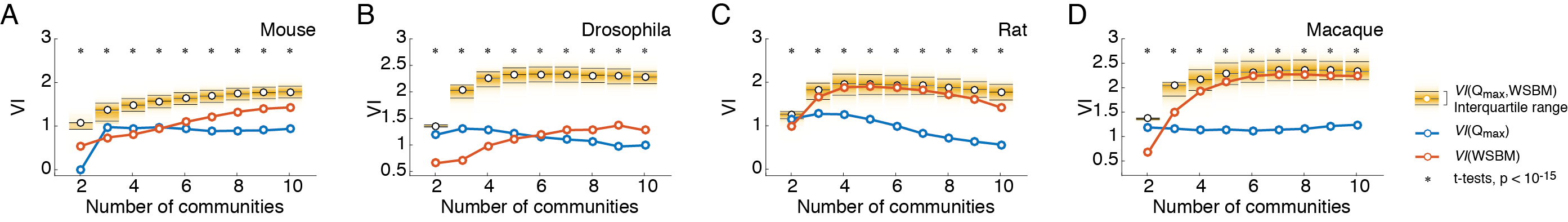}}
		\caption{\textbf{Partition dissimilarity for non-human connectome data.} We show mean within-technique partition dissimilarity (variation of information; VI) in blue and orange for $Q_{max}$ and the WSBM, respectively, along with the between-technique dissimilarity (yellow). Panels (\emph{A}-\emph{D}) depict VI as a function of the number of communities for mouse, \emph{Drosophila}, rat, and macaque, respectively.} \label{communitySimilarity4}
	\end{center}
\end{figure*}

\subsection*{Comparing the WSBM with $Q_{max}$}

As in the main text, we compared the partitions detected using the WSBM with those detected using $Q_{max}$. This involved a partition-level comparison using variation of information (VI), a community-level comparison using the community assortativity measure, $\mathcal{A}_r$, a regional-level analysis based on a regional assortativity score, $a_i$, and functional data analysis.

\begin{figure*}[t]
	\begin{center}
		\centerline{\includegraphics[width=1\textwidth]{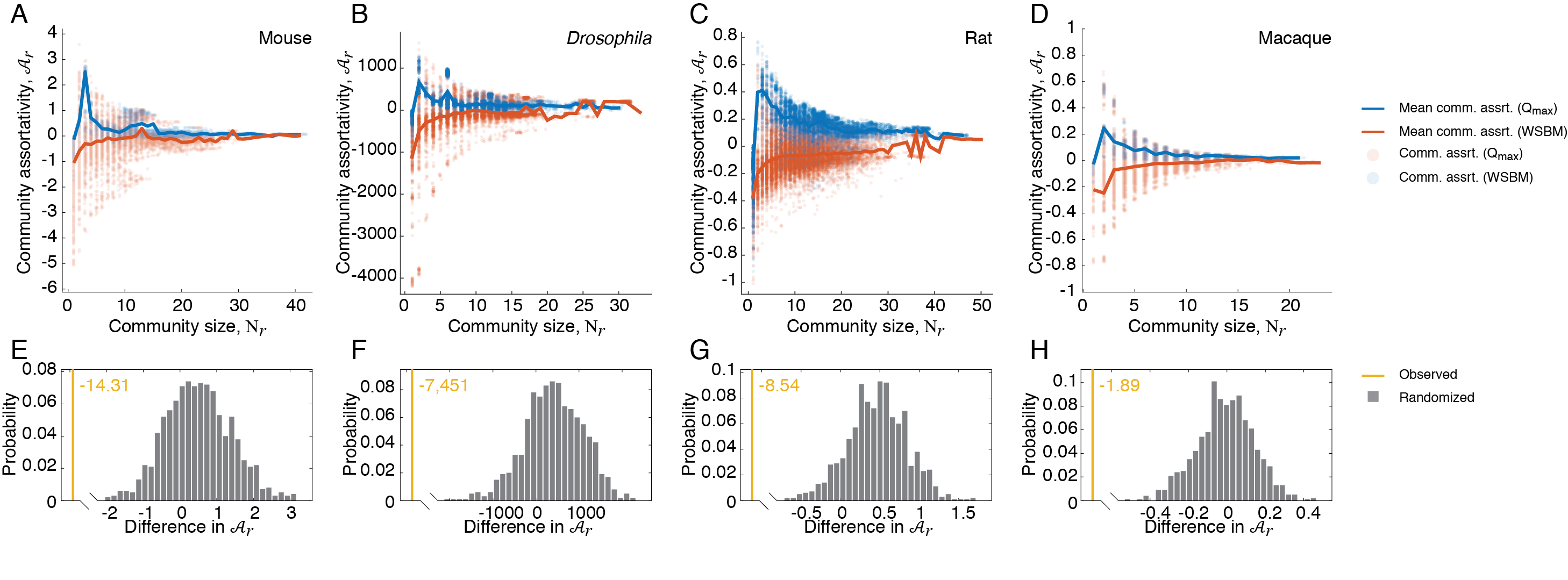}}
		\caption{\textbf{Community assortativity for non-human connectome data.} We show the assortativity of each detected community for both the WSBM (orange) and $Q_{max}$ (blue) as a function of community size. The solid curves represent the mean community assortativity for each technique as a function of community size. Panels (\emph{A}-\emph{D}) depict scores for mouse, \emph{Drosophila}, rat, and macaque, respectively. The accompanying panels (\emph{E} - \emph{H}) are the results of the functional data analysis, in which we compared the mean community assortativity curves. Yellow lines represent the observed test statistic; gray bars represent the null distribution.} \label{communityAssortativity4}
	\end{center}
\end{figure*}

\begin{figure*}[t]
	\begin{center}
		\centerline{\includegraphics[width=0.75\textwidth]{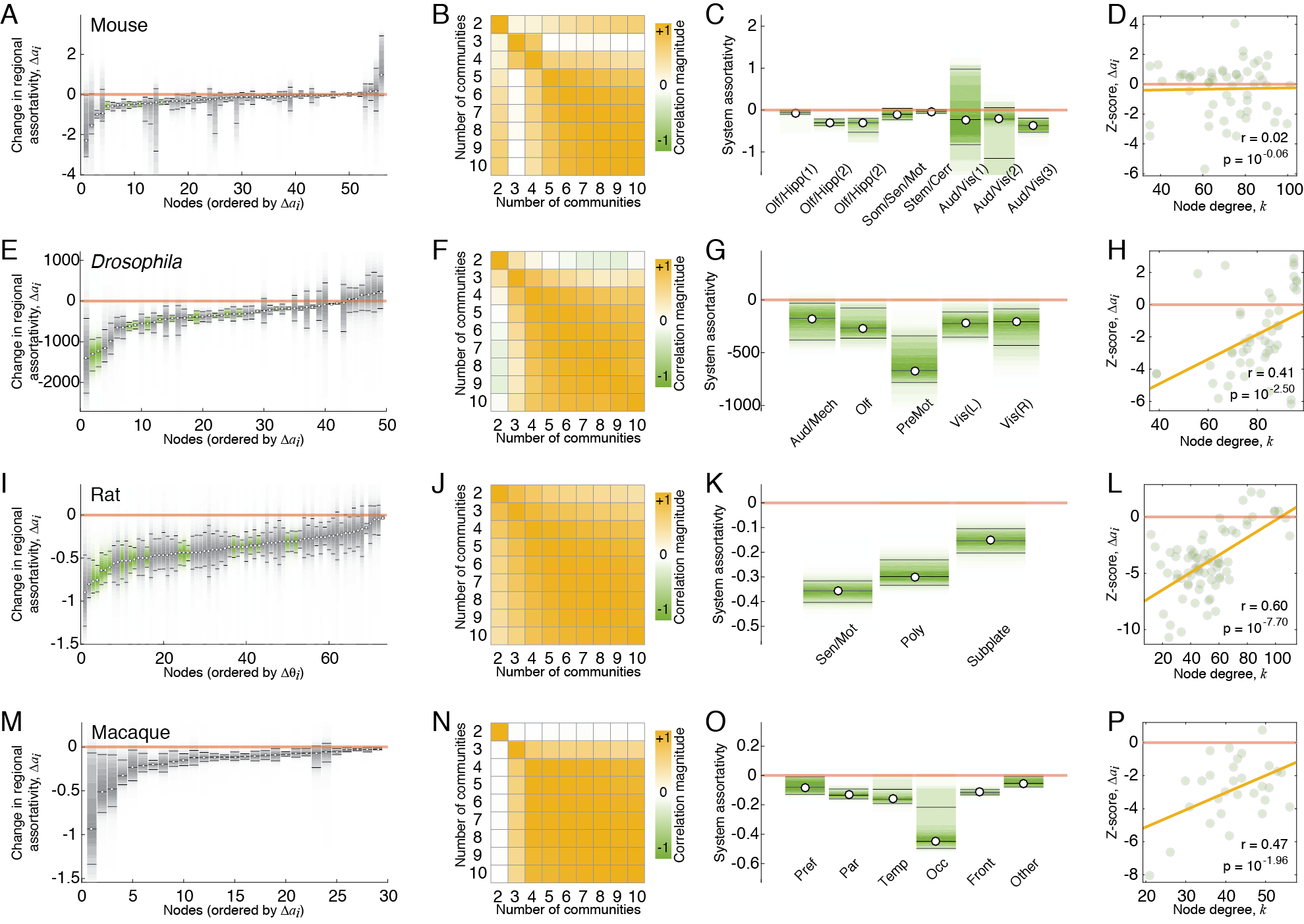}}
		\caption{\textbf{Regional assortativity for non-human connectome data.} Panels (\emph{A},\emph{E},\emph{I},\emph{M}) depict differences in regional assortativity scores, $\Delta a_i$, for mouse, \emph{Drosphila}, rat, and macaque. Panels (\emph{B},\emph{F},\emph{J},\emph{N}) show the similarity (correlation) of changes in regional assortativity as a function of the number of communities. Panels (\emph{C},\emph{G},\emph{K},\emph{O}) show the same regional assortativity changes grouped according to functional systems for each connectome dataset. Panels (\emph{D},\emph{H},\emph{L},\emph{P}) show the relationship of regional assortativity to node degree.} \label{regionalAssortativity2NonHuman}
	\end{center}
\end{figure*}

For the partition-level analyses, we observed that the partitions detected using either the WSBM or $Q_{max}$ were more internally consistent than they were similar to one another as we varied the number of communities from $K = 2$ to $K = 10$ (Fig.~\ref{communitySimilarity4}A-E) ($t$-tests, $p < 10^{-15}$). We found similarly consistent results when we examined community assortativity (Fig.~\ref{communityAssortativity4}A-H), with $Q_{max}$ communities being more assortative than WSBM communities (controlling for size, $p < 10^{-3}$), and when we examined regional assortativity (Fig.~\ref{regionalAssortativity2NonHuman}), with the majority of regions exhibiting decrements in regional assortativity under the WSBM.

\subsection*{Maximally assortative set}
\begin{figure*}[t]
	\begin{center}
		\centerline{\includegraphics[width=0.75\textwidth]{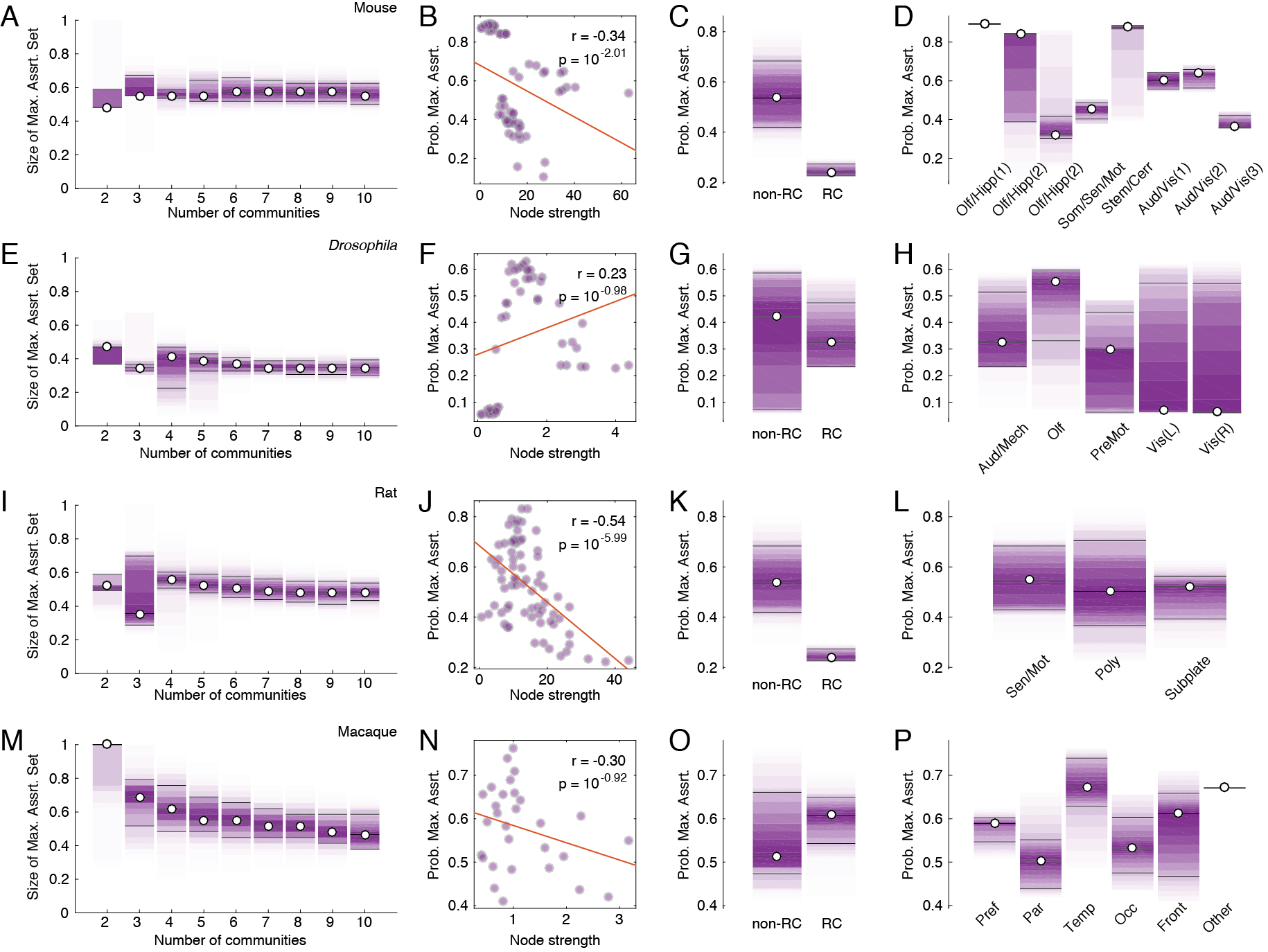}}
		\caption{\textbf{Maximaly assortative set for non-human connectome data.} We show the size of the maximally assortative set (as a fraction of total number of nodes) for the (\emph{A}) mouse, (\emph{E}) \emph{Drosophila}, (\emph{I}) Rat, and (\emph{M}) macaque connectomes. For each connectome dataset, we also plot the probability that a node gets assigned to the maximally assortative set as a function of the logarithm of node strength (\emph{B,F,J,N}), whether or not (\emph{C,G,K,O}), and according to the functional system to which each node was assigned (\emph{D,H,L,P}).} \label{maximallyAssortativeSetNonHuman}
	\end{center}
\end{figure*}

In the main text, we demonstrated that there was a subset of around 75\% of all network nodes whose communities were all mutually assortative -- i.e. the minimum within-community connection density of all communities exceeded the maximum between-community density of any pair of communities. Here, we reproduce this analysis for the non-human connectome data. In general, these results corroborate those of the main text. In particular, we find that the maximally assortative set is never empty (Fig.~\ref{maximallyAssortativeSetNonHuman}A, E, I, M), though its size varies considerably across networks, with an average size of $0.37 \pm 0.11$ in \emph{Drosophila} up to $0.58 \pm 0.09$ in mouse.

We also observed that, in general, participation in the maximally assortative set was negatively correlated with node strength (though we observe an opposite trend in the \emph{Drosophila} dataset; Fig.~\ref{maximallyAssortativeSetNonHuman}F). Similarly, in terms of rich club participation in the maximally assortative set, we observed that rich club nodes were less likely to participate in the maximally assortative set than non-rich club nodes (though again, we see an opposite effect in the macaque dataset). For completeness, we also show participation in the maximally assortative set by functional system.

\subsection*{Community morphospace analysis}
\begin{figure*}[t]
	\begin{center}
		\centerline{\includegraphics[width=0.75\textwidth]{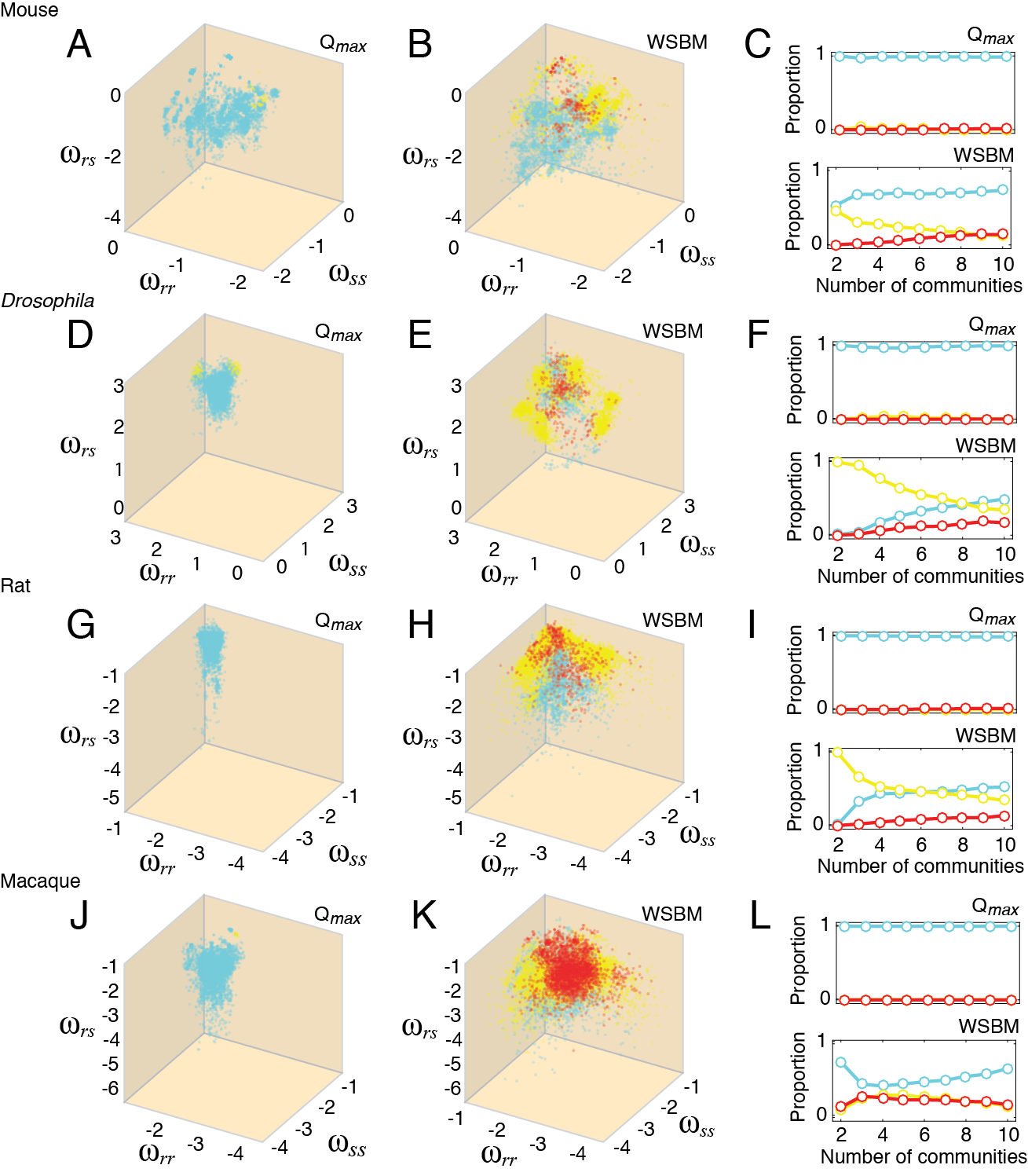}}
		\caption{\textbf{Community morphospace analysis of non-human connectome data.} Rows 1-4 depict mouse, \emph{Drosophila}, rat, and macaque data. Panels (\emph{A},\emph{D},\emph{G},\emph{J}) Community morphospaces constructed based on $Q_{max}$ partitions. Panels (\emph{B},\emph{E},\emph{H},\emph{K}) Community morphospaces constructed based on WSBM partitions. Panels (\emph{C},\emph{F},\emph{I},\emph{L}) Motif proportions for both $Q_{max}$ (\emph{top}) and the WSBM (\emph{bottom}) as a function of the number of detected communities.} \label{communityMorphospaceWeighted_nonHuman}
	\end{center}
\end{figure*}

As with the human connectome data presented in the main text, we compared the morphospaces populated by communities detected using the WSBM and $Q_{max}$ (Fig.~\ref{communityMorphospaceWeighted_nonHuman}). We repeated this analysis separately for each non-human connectome dataset. Unlike the human connectome data, which was directed (and so the following equality holds: $\omega_{rs} = \omega_{sr}$), the directed non-human networks allowed us to examine a greater number of community interactions. Nonetheless, the results of these analyses were entirely consistent with those described in the main text using the human connectome data. Namely, we observed that for all non-human data and over the range $K = 2, \ldots , 10$, $Q_{max}$ discovers almost exclusively assortative community motifs. The WSBM, on the other hand, detects disproportionately more disassortative and core-periphery motifs than $Q_{max}$. As before, we confirm these results using functional data analysis, demonstrating the difference in motif proportions were unexpected under a permutation-based null model (mouse, $p = 0.003$; \emph{Drosophila}, $p < 10^{-3}$; rat, $p < 10^{-3}$; macaque, $p < 10^{-3}$).

\begin{figure*}[t]
	\begin{center}
		\centerline{\includegraphics[width=0.75\textwidth]{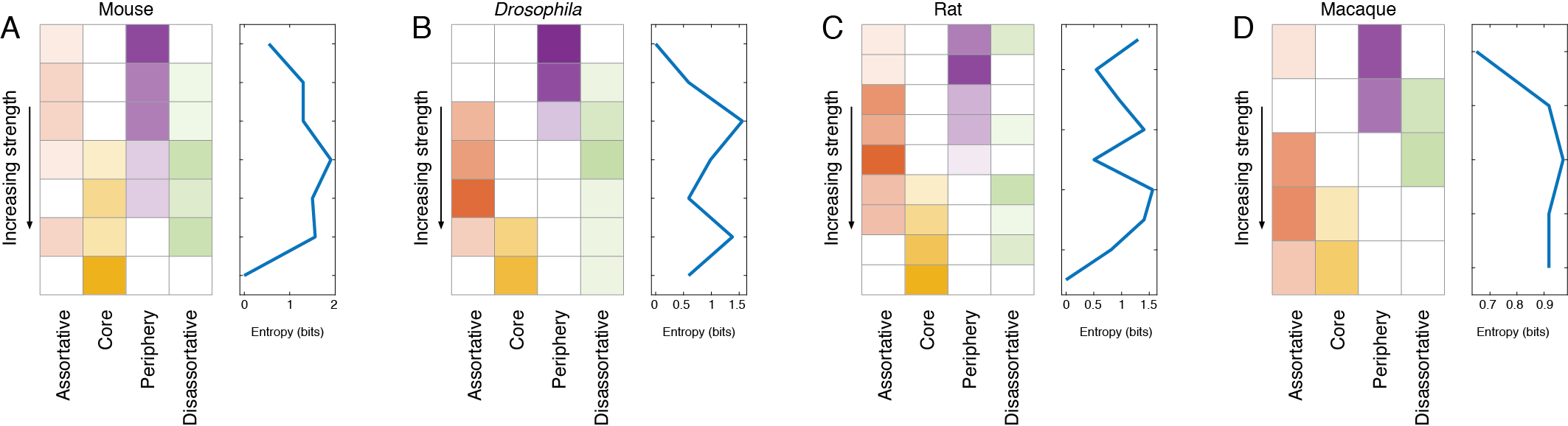}}
		\caption{\textbf{Relative proportion of community motifs as a function of node strength.} (\emph{A}) Mouse, (\emph{B}) \emph{Drosophila}, (\emph{C}) rat, and (\emph{D}) macaque. For each connectome dataset, we show the relative proportion of each motif type (\emph{left}) and the entropy within each bin (\emph{right}).} \label{communityParticipationAllSpecies}
	\end{center}
\end{figure*}

Based on the results of the community morphospace analysis, we were able to assign individual network nodes to the motif class that they expressed differentially, assign them to bins based on their node strength, and compute the entropy of the motif class distribution within each bin (Fig.~\ref{communityParticipationAllSpecies}). As in the main text, we observed that entropy peaked for middle-strength nodes, while the lowest- and highest-strength nodes were associated with low entropies, suggesting that they participate in a much narrower range of classes. Also, as described in the main text, low-strength bins were dominated by assortative and periphery motifs while high-strength bins were dominated by core motifs.

These supplementary results, in addition to those described in the previous section, bolster the findings from the main text, and suggest that our results cannot be explained by shortcomings of any one connectome reconstruction technique. As before, these supplementary results demonstrate that connectome data can be described with a rich meso-scale architecture that is detectable with the WSBM.

\subsection*{Functional relevance of WSBM communities for mouse connectome}
With the human data, we were able to demonstrate the functional relevance of the WSBM communities by comparing the within- and between-community densities of functional connections, which were calculated given a group-representative resting-state FC matrix. Unfortunately, we do not have comparable functional recordings for any of the non-human datasets. The mouse dataset, however, does include regional gene expression profiles and the similarity of two regions' profiles is often taken to indicate the extent to which those regions participate in a shared set of functions \cite{lamb2006connectivity, subramanian2005gene}.

\begin{figure*}[t]
	\begin{center}
		\centerline{\includegraphics[width=0.85\textwidth]{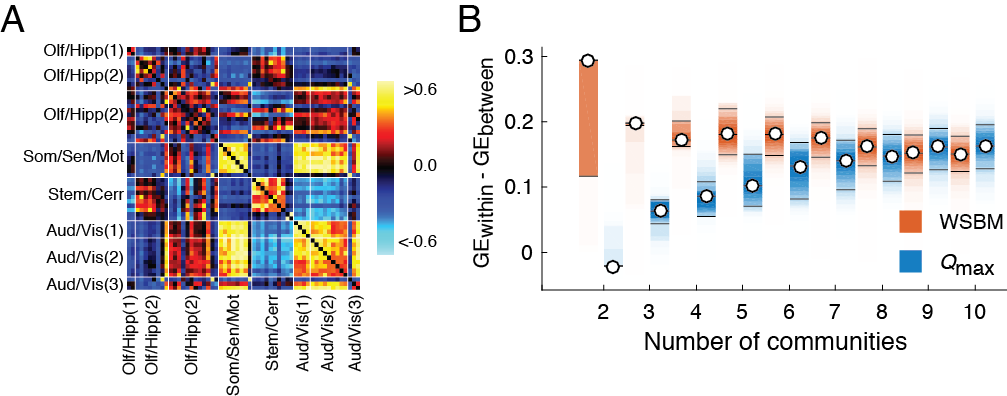}}
		\caption{\textbf{Correlated gene expression profiles and the WSBM in mouse data.} (\emph{A}) Correlation matrix of gene expression profiles. (\emph{B}). Difference in within- \emph{versus} between-community correlations for both the WSBM (orange) and $Q_{max}$ (blue).} \label{geneMat}
	\end{center}
\end{figure*}

With this assumption in mind, we generate a region-by-region matrix of gene expression correlation coefficients, which we can view as analogous to the human FC matrix. We can then impose partitions detected using the WSBM and $Q_{max}$ onto this matrix and compute the mean within- and between-community correlation. As before, the larger the difference in these two variables, the better job a partition does in uncovering functionally meaningful communities. We computed the difference for all partitions detected from $K = 2$ to $K = 10$. We observed that over the sub-range $K = 2$ to $K = 8$, the WSBM partitions resulted in greater differences between within- and between-community correlations (one-tailed $t$-test; $p < 10^{-5}$). This suggests that over that same range, the WSBM detects communities of brain regions that are correlated with one another but, in general, anti-correlated between communities. Moreover, the WSBM performs better than $Q_{max}$ (Fig.~\ref{geneMat}), indicating that on average the WSBM detects communities that are better aligned with the correlated expression profiles of brain regions in the mouse than the communities detected using $Q_{max}$.

\end{document}